\documentclass[12pt]{article}
\usepackage{caption}
\usepackage{subcaption}
\usepackage{geometry}
\usepackage{titlesec}
\usepackage{multirow}
\usepackage[T1]{fontenc}
\usepackage[utf8]{inputenc}
\usepackage{authblk}
\usepackage{graphics}
\usepackage{makecell}
\usepackage{float}
\usepackage{bigints} 
\usepackage[dvipsnames]{xcolor}
\providecommand{\keywords}[1]{\textbf{Keywords: } #1}
\titleformat{\section}{\centering\large\scshape}{\thesection}{1em}{}
\titleformat{\subsection}{\centering\normalsize\scshape}{\thesubsection}{1em}{}

\usepackage{hyperref}
\hypersetup{
	colorlinks=true,
	linkcolor=blue,
	filecolor=blue,      
	urlcolor=blue,
	citecolor=blue
}

\usepackage{natbib}
\usepackage{graphicx} 
\graphicspath{{art/}}

\usepackage{tikz}
\usetikzlibrary{fit,positioning,arrows,automata,calc}
\tikzset{
	main/.style={circle, minimum size = 5mm, thick, draw =black!80, node distance = 10mm},
	connect/.style={-latex, thick},
	box/.style={rectangle, draw=black!100}
}
\usetikzlibrary{decorations.pathreplacing}

\usepackage{setspace}
\usepackage{xr}
\usepackage{color}
\usepackage{amsmath, amssymb}
\usepackage{amsfonts,dsfont,mathrsfs,mathtools}
\usepackage{booktabs,multirow,array}
\usepackage{bm}
\usepackage{arydshln}
\usepackage[vlined,linesnumbered,ruled,resetcount]{algorithm2e}

\newcommand{\R}{\mathds{R}}

\newcommand{\D}{{\rm d}}

\newcommand{\Law}{\mathcal{L}}

\def\simiid{\stackrel{\mbox{\scriptsize{iid}}}{\sim}}

\allowdisplaybreaks[3]

\newcommand{\eX}{X^{(\varepsilon)}}
\newcommand{\eA}{A^{(\varepsilon)}}

\newcommand{\eM}{M^{(\varepsilon)}}
\newcommand{\indicator}[2]{\mathds{1}_{#1}\left(#2\right)}
\newcommand{\norm}[2]{\lvert\lvert #1 \rvert\vert_{#2}}

\begin{document}

\def\spacingset#1{\renewcommand{\baselinestretch}%
{#1}\small\normalsize} \spacingset{1}


 \title{\scshape\LARGE{ Model-based clustering of time-dependent observations with common structural changes}}
 
 \author[1]{Riccardo Corradin}
 \author[1]{Luca Danese}
 \author[2]{Wasiur R. KhudaBukhsh}
 \author[1]{Andrea Ongaro}
 \affil[1]{\normalsize{Department of Economics, Management and Statistics, University of Milano-Bicocca, Milano, Italy}}
 \affil[2]{\normalsize{School of Mathematical Sciences, University of Nottingham, Nottingham, United Kingdom}}
 
 \maketitle
 \date{}

\bigskip
\begin{abstract}
	
\noindent
We propose a novel model-based clustering approach for samples of time series. We assume as a unique commonality that two observations belong to the same group if structural changes in their behaviours happen at the same time. We resort to a latent representation of structural changes in each time series based on random orders to induce ties among different observations. Such an approach results in a general modeling strategy and can be combined with many time-dependent models known in the literature. Our studies have been motivated by an epidemiological problem, where we want to provide clusters of different countries of the European Union, where two countries belong to the same cluster if the spreading processes of the COVID-19 virus had structural changes at the same time.

\vspace{12pt}

\noindent\keywords{Change points; COVID-19; Model-based clustering; Time series.}
\end{abstract}
  
\section{Introduction}

Detecting patterns and similarities between observations is a key topic in modern statistics, especially in large datasets from which we can extract information. Among fundamental approaches, cluster analysis comprises a collection of statistical tools to gather insights from a set of data by looking for groups with similar features, defining some homogeneous subsets of observations. In particular, in a model-based clustering setting data are described by a mixture model, where two observations belong to the same cluster if they are assigned to the same component of the mixture. Equivalently, each observed datum is associated with a latent parameter, which indexes a specific mixture component, and two observations are clustered together if their latent parameters have the same value. 
Here we consider a scenario involving multiple time series, with each observation representing a series of values observed over time. 
We introduce a new method for clustering these time series based on specific characteristics, such as shared structural changes over time.


While various methods exist for modeling time-dependent data, a desirable property is the ability to account for shocks in the observed values. These alterations occur when an external event significantly impacts the phenomena being analyzed, altering its future behavior. 
The identification of these points, and the study of models accommodating those structural changes is usually termed \textit{change points detection}. Seminal contributions on this topic are \citet{Pag54,Pag57}, who employed a frequentist approach using hypothesis tests to identify a single structural change in a time-dependent parameter. Later, such approaches have been extended in different directions by considering different methods to detect a change point. These methods include parametric and nonparametric approaches and cases where the position of the change point is assumed to be known or unknown \cite[see, e.g.,]{brodsky1993nonparametric}. However, trying to detect a single change point might be too restrictive since in many situations more than one change may occur. Different works then extended these methods to the problem of multiple change points detection. \cite{NiuHangZhang2016} provided a general review of multiple change points detection methods for mean changes in Gaussian model. 
\cite{InclanTiao1994} proposed an iterative method to detect changes in the variance of a time series, where a change point is detected whenever an abrupt change occurs in the cumulative sum of squares of the time series realisations. Similarly, \cite{ChenGupta1997} proposed a method to infer the structural changes by selecting the configuration of change points that minimises the Schwarz information criterion. See \cite{ChenJieGuptaArjun2021} for a review on models for change point models and possible extensions.

Beside frequentist approaches, many contributions have been introduced also in the Bayesian literature. Early studies rely on an hypothesis tests construction to detect a single structural change, such as \citet{Cher64}. Recent contributions investigated different strategies to accommodate for multiple change points. Among these, \citet{Bar92, Bar93} considered a formulation of the problem that is based on a Product Partition Model \citep[PPM,][]{Har90} construction to detect mean changes in an observed time series, which is assumed to be Gaussian distributed. The underlying model is based on a latent random partition of the parameters, where two parameters belong to the same block of the partition if they share the same value, and two distinct blocks of the partition describe two different regimes of the observed quantities. Later this approach has been extended in many directions. \citet{Los02} proposed a strategy to detect changes in both mean and variance of a Gaussian kernel distribution over time, based on the PPM construction. \citet{Qui03} studied the connection between PPM and the Dirichlet process in the context of change points. 
\citet{Fue10} proposed a similar approach, but considering as partition model the Exchangeable Partition Probability Function (EPPF) of a Dirichlet process and restricting the support of such EPPF to the random orders space by assigning probability zero to partitions that do not preserve the order. Later \citet{Mar14} considered a transformation of the EPPF arising from a Pitman-Yor process \citep{Pit95} that preserves the properties of the EPPF, such as the symmetry of the corresponding probability distribution and the distribution of the number of distinct blocks, but restricts the support of the function to the space of random orders, following the pioneering studies of~\citet{Pit06}. More recent contributions considered multivariate cases \citep{corradin2022}, correlated change points for multiple time-dependent observations \citep{Quinlan-2022}, different change points for different parameters \citep{Pedroso-2021} and models where change points detection is performed assuming a temporal dependency among random partitions \citep{Page2022,giampino2024localleveldynamicrandom}.

We introduce an approach that employs a model-based clustering framework to group observations exhibiting common structural changes, while other individual parameters and regimes are observation-specific. Our goal is to cluster observations that experience a shock simultaneously, even if their behaviors differ. 
To our knowledge, no methods exist for model-based clustering of time series that primarily focus on change points. However, there are various contributions aimed at clustering multiple time series. For a comprehensive review of different clustering techniques for time series see \cite{AGHABOZORGI201516}. In a Bayesian framework, early contributions to time series model-based clustering can be found in \cite{Frühwirth-Schnatter-2008}, who examined a scenario with a fixed number of clusters, grouping observations based on latent parameters that define their behaviors. 
Later extensions include different kernel functions \citep{Juarez-Steel-2010} and categorical data \citep{Frühwirth-Schnatter-Pamminger-2008}, among others. An allied approach to our contribution can be found in \citet{same-2011}, where the authors propose a model to cluster together time series with changes in their regimes, resorting to polynomial representations of the time-dependent observations whose degrees can change over time.

Our primary motivations behind this study comes from the recent COVID-19 pandemic. Epidemic models such as the Kermack--McKendrick renewal equation models \citep{Kermack1927} and their stochastic analogues have been studied for many decades now \citep{AnderssonBritton2000}. Along with their study, parameter inference methods have been also developed \citep[see, e.g,][]{Becker1993Martingale,Becker1993Parametric,Choi2011Inference,Fierro2015Statistical,Kypraios2017tutorial}. See \cite{Britton2019Stochastic} for an overview on this topic. Due to the recent COVID-19 pandemic, these methods received a renewed interest.  
For example, frequentist methods \citep{Bong2023Frequentist} and other techniques for calibration of computer models \citep{Fadikar2018Calibrating,Baker2022Analyzing} have been investigated successfully in the last years. Similarly, the problem of parameter inference from a Bayesian perspective was discussed for mechanistic \citep{Seymour2022Bayesian,Chitwood2022Reconstructing,khudabukhsh2022projecting} and semi-mechanistic models of COVID-19 \citep{Bhatt2023Semimechanistic}.
More recent contributions 
accommodate for different behaviour of the model over time. For example, \cite{Hong2020Estimation}, \cite{Gleeson2021Calibrating} and \cite{Wascher2024IDSA} discussed models with time-varying infection rates, while structural changes in such parameters were modeled in \cite{Hua24}. Here, we want to cluster different countries on the basis of their structural changes in the dynamics dictated by 
a stochastic epidemic model.

The paper is structured as follows. Section \ref{sec:model} introduces a model to cluster together time-dependent observations with common change points. Section \ref{sec:posterior} describes the distributions of interest that can be used to construct a sampling strategy. In Section \ref{sec:algorithm} we present a Markov chain Monte Carlo (MCMC) procedure to sample realizations from the posterior distribution of interest. Section \ref{sec:epi} shows the epidemiological study which motivates our contribution. Conclusions and future developments are discussed in Section \ref{sec:discussions}. Further details on algorithm implementations, illustrations, and the epidemiological model are deferred to the Web Appendix. Finally, an implementation of the methodologies presented in the manuscript is available in the \cite{lucarepository} repository.

\section{Modeling multiple time series with common change points} \label{sec:model}


We introduce a novel approach to cluster together time series with respect to common structural changes, and no further local similarities are assumed among them. In this sense, observations might be locally indexed by different subject-specific behaviours, but cluster together if they change regime on the same time instants. Let $\bm y_i = \{y_{i1}, \dots, y_{iT}\}$, be the generic $i$th time series we observe, where $y_{i,t}$ is a random quantity taking values on a space $\mathbb Y$, which depends on the specific quantity we observe, and $\mathcal Y = \{\bm y_1, \dots, \bm y_n\}$ denotes the whole sample. For simplicity, here we assume that time series are observed at the same discrete times $\{1, \dots, T\}$, however our approach works even in more general scenarios. See, e.g., Section~\ref{sec:epi} where we consider a model tailored for epidemiological spreads as kernel function.

Regarding the distribution of each time series, we assume that our data are sharing the same model structure over time with Markovian dependence, where $\Law(y_{i,t} \mid y_{i,t-1}, \theta_{i,t}), \, t = 1, \dots, T$ denotes the distributional law of $y_{i,t}$ given the previous observation $y_{i,t-1}$ and the local parameter $\theta_{i,t}$, with $\theta_{i,t} \in \Theta$ for all $i = 1, \dots, n$ and $t = 1, \dots, T$. We remark that the modeling structure presented in the manuscript can be extended to cases with longer memories by substituting the distribution $\Law$ with suitable alternatives. The local behaviour of the model for the $i$th observation is dictated by the sequence of latent parameters $\theta_{i,1}, \dots, \theta_{i,T}$, which are assumed to have ties over time. A change point for the $i$th time series occurs when the parameter at time $t$ differs from the previous value, i.e. $\theta_{i,t} \neq \theta_{i,t-1}$. Equivalently, each observation is divided into one or more blocks, where realisations within each block share the same value of $\theta_{i,t}$. When the value of the parameter changes a change point occurs and a new block begins. The distribution of the sequence of parameters $\theta_{i,1}, \dots, \theta_{i, T}$ can be factorised in two independent terms
\begin{equation}\label{eq:fact_dist}
	\Law(\theta_{i,1}, \dots, \theta_{i,T}) = \Law(\rho_i) \Law(\theta_{i,1}^*, \dots, \theta_{i,m_i}^*).
\end{equation}
The first term $\Law(\rho_i)$ denotes the distribution of the random order $\rho_i$ inducing changes in the sequence of parameters $\theta_{i,1}, \dots, \theta_{i,T}$. Here, by random order $\rho_i$ we mean a random partitioning of $\{1, \dots, T\}$ into $m_i$ blocks $A_{i,1}, \dots, A_{i, m_i}$ satisfying the following properties 
\begin{itemize}
	\item[-] $A_{i,\ell} \cap A_{i, j} = \emptyset$ for all $\ell, j \in \{1, \dots, m_i\}$ with $\ell \neq j$;
	\item[-] $A_{i,1} \cup \cdots \cup A_{i,m_i} = \{1, \dots, T\}$;
	\item[-] For any $j, \ell \in \{1, \dots, m_i\}$ with $j < \ell$, $w \in A_{i,j}$ and $s \in A_{i,\ell}$, we have $w < s$. 
\end{itemize}
The first two conditions are properties of a generic partition of $T$ elements, while the latter restricts the partition space to the subset satisfying the ordering constraint. The second term in~\eqref{eq:fact_dist} denotes the distribution of $\theta_{i,1}^*, \dots, \theta_{i,m_i}^*$, the unique values out of $\theta_{i,1}, \dots, \theta_{i,T}$, whereas the generic $\theta_{i,j}^*$ determines a common local behaviour and it is shared by all the observations whose times are allocated in the $j$th block of $\rho_i$. Hence, a time series switches its regime when it moves from a block to another of $\rho_i = \{ A_{i,1}, \dots, A_{i, m_i}\}$, while the local behaviours are fully characterised by $\theta_{i,1}^*, \dots, \theta_{i,m_i}^*$.

\tikzset{rectangle/.append style={draw=black}}
\begin{figure}[!ht]
	\includegraphics[width=\textwidth]{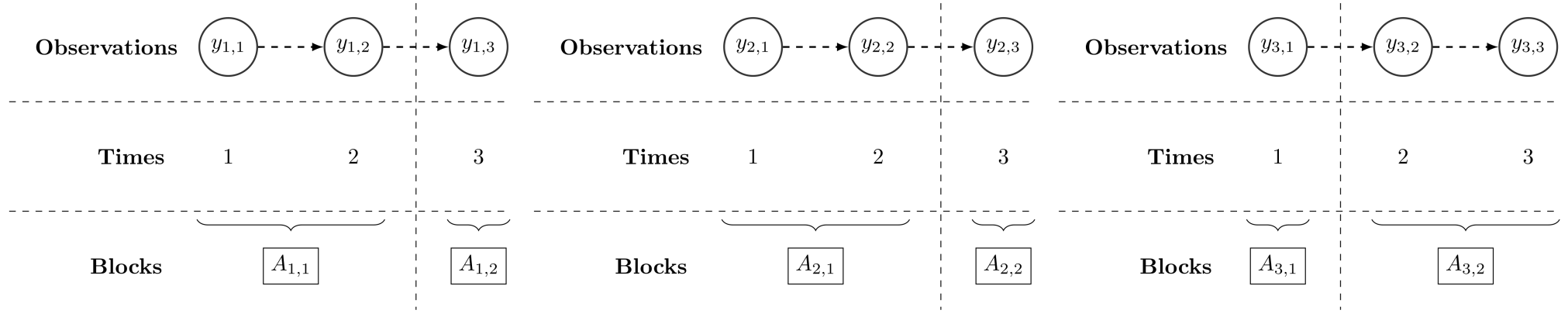}
	\caption{Graphical representation of three time series $\bm y_1, \bm y_2, \bm y_3$, where each series is observed at $T=3$ times, with their corresponding latent partition, whereas $\bm y_1$ and $\bm y_2$ are in the same cluster while $\bm y_3$ is in a separate cluster.}\label{fig:graph_ts}
	\label{fig:sketch_cps}
\end{figure}

Our aim is to cluster together time series in homogeneous groups sharing the same change points. To this end, we assume the latent orders $\mathcal R = \{\rho_1, \dots, \rho_n\}$ distributed independently according to a discrete distribution $\tilde p$, with
\begin{equation}\label{eq:disc_prior}
	\tilde p(\rho) = \sum_{r=1}^{2^{T-1}}\pi_r\delta_{ \tilde \rho_r}(\rho),
\end{equation}
where $\{\tilde\rho_1, \dots, \tilde\rho_{2^{T-1}}\}$ is an exhaustive collection of all the possible orders of $T$ elements, $(\pi_1, \dots, \pi_{2^{T-1}})$ is a vector of probabilities taking values in the $(2^{T-1}-1)$-dimensional simplex space, and $\delta_a(\cdot)$ denotes a Dirac measure at $a$.  
The discreteness of \eqref{eq:disc_prior} implies a positive probability of having ties in $\mathcal R$. Ties represent clusters of observations sharing the same change points, i.e. changing their local behaviour at the same time. The observations are then divided according to a partition $\lambda =\{ B_1, \dots, B_k\}$ of the indices $\{1, \dots, n\}$, for which $B_\ell \cap B_j = \emptyset$ for all $\ell, j \in \{1, \dots, k\}$ with $\ell \neq j$, and $B_1 \cup \cdots \cup B_k = \{1, \dots, n\}$. Figure~\ref{fig:sketch_cps} shows an example with $n = 3$ time series observed for $T = 3$ time instants with a single change point. The left and middle series belong to the same cluster, as the change point occurs at $t=3$, while the right one forms a stand-alone cluster with a change point at time $t = 2$.

The model specification is completed by setting a distribution for the weights $\pi_1, \dots, \pi_{2^{T-1}}$, in the following sections assumed to be Dirichlet distributed, by choosing a distribution for the observed data $\Law(y_{i,t}\mid y_{i,t-1}, \theta_{i,t})$ and by setting a prior distribution for the parameters $\theta_{i,t}$s, here denoted by $P_0$. We remark that, since each observed time series is associated with a latent order, its likelihood contribution can be factorised as a product of terms sharing the same local behavior. The model we consider can be then written in its hierarchical form as
\begin{equation}\label{mod:spec}
	\begin{split}
		\bm y_i \mid \rho_i, \bm \theta_i^* &\sim \prod_{j=1}^{m_i} \prod_{t = t_{i,j}^-}^{t_{i,j}^+} \Law(y_{i,t}\mid y_{i,t-1}, \theta_{i,j}^*),\quad i = 1, \dots, n,\\
		\rho_i \mid \tilde p(\rho)&\simiid \tilde p(\rho), \quad i = 1, \dots, n,\\
		(\pi_1,\dots,\pi_{2^{T-1}}) &\sim \textsc{Dir}(\alpha_1, \dots, \alpha_{2^{T-1}}),\\
		\theta_{i,j}^*&\simiid P_0(\theta),\quad j = 1, \dots, m_i, \; i = 1, \dots, n,
	\end{split}
\end{equation}
where $t_{i,j}^- = \min\left\{t: t \in A_{i,j} \right\}$ denotes the first time index belonging the $j$th block of the latent order associated with the $i$th observation, $t_{i,j}^+ = \max\left\{t: t \in A_{i,j} \right\}$ the last time index, and with the proviso that the first time of a new regime in the observed quantities does not depend on observed values in past regimes, i.e.  $\Law(y_{i,t_{i,j}^-} \mid y_{i,t_{i,j}^- - 1}, \theta_{i,j}^*) = \Law(y_{i,t_{i,j}^-} \mid \theta_{i,j}^*)$. 

Specific distributional choices of $\Law(y_{i,t}\mid y_{i,t-1},\theta_{i,t})$ depend on the data we are considering. For example, in the following sections we present an illustration where the data are real-valued quantities modeled through an univariate Ornstein-Uhlenbeck process (Section~\ref{subsec:application_time_series} of Web Appendix) and a case with survival data arising from a susceptible-infected-removed model (Section~\ref{sec:epi}). However, any time-dependent model with discrete or discretisable observational time can be embedded in the specification of a model as in~\eqref{mod:spec}. The distribution of the parameters $\mathcal T^* = \{\bm \theta_1^*, \dots, \bm\theta_n^*\}$ depends on the distributional assumption of $\Law(y_{i,t}\mid y_{i,t-1},\theta_{i,t})$. 
Futher, $\mathcal T^*$ can be marginalised in case we are not interested on posterior inference for local behaviours, but only on common structural change times. 

\section{Posterior distribution of interest} \label{sec:posterior}
Our main interest lies in the latent partition defining groups of time series which share the same change points. The posterior distribution of such quantity is not available in a closed form, but we can identify a tractable expression from which we can produce a sample. Hence, we can start from the joint distribution of the observed time series $\mathcal Y$, latent orders $\mathcal R$, unique values of the parameters $\mathcal T^*$ and the vector of probabilities $\bm \pi = (\pi_1, \dots, \pi_{2^{T-1}})$. The distribution of the previous quantities can be then expressed, in force of the chain rule, as product of the corresponding conditional distributions, with 
\begin{equation}\label{eq:mod_prod}
	\Law(\mathcal Y, \mathcal T^*, \mathcal R, \bm \pi) = \Law(\mathcal Y \mid \mathcal T^*, \mathcal R) \Law(\mathcal T^* \mid \mathcal R) \Law(\mathcal R \mid \bm \pi) \Law(\bm \pi),
\end{equation}
where $\mathcal Y$ is independent of $\bm \pi$ conditioned on $\mathcal T^*$, $\mathcal R$ and $\mathcal T^*$ is also independent of $\bm \pi$ conditioned on $\mathcal R$. We can make the joint distribution in~\eqref{eq:mod_prod} explicit by using the distributions in~\eqref{mod:spec}, with the likelihood term, here denoted by $\Law(\mathcal Y \mid \mathcal T^*, \mathcal R)$, being the product of the data distribution for all the observed time series. We then have
\begin{equation}\label{eq:model_expanded}
	\Law(\mathcal Y, \mathcal T^*, \mathcal R, \bm \pi) = \prod_{i=1}^n\prod_{j=1}^{m_i} \prod_{t = t_{i,j}^-}^{t_{i,j}^+} \Law(y_{i,t}\mid y_{i,t-1}, \theta_{i,j}^*) \prod_{i=1}^n \prod_{j=1}^{m_i} P_0(\theta_{i,j}^*) \prod_{i=1}^n \tilde p (\rho_i) \frac{1}{\mathrm B(\bm \alpha)}\prod_{r = 1}^{2^{T-1}} \pi_r^{\alpha_r - 1},
\end{equation}
where $B(\bm \alpha)$ denotes the multivariate beta function of parameter $\bm \alpha$.

We are interested in partitioning the observations on the base of their latent orders. Therefore, the unique values of the parameters $\mathcal T^*$ can be marginalised from the previous equation, obtaining the following expression
\begin{equation*}
	\begin{split}
		\Law(\mathcal Y, \mathcal R, \bm \pi) &= \prod_{i=1}^n\prod_{j=1}^{m_i}\mathcal M(\{y_{i,t} \mid t \in A_{i,j}\}) \prod_{i=1}^n \sum_{r=1}^{2^{T-1}} \pi_r\delta_{\tilde\rho_{r}}(\rho_i) \frac{1}{\mathrm B(\bm \alpha)}\prod_{r = 1}^{2^{T-1}} \pi_r^{\alpha_r - 1},
	\end{split}
\end{equation*}
whereas $\mathcal M(\{y_{i,t} \mid t \in A_{i,j}\})$ denotes the marginal distribution of the $i$th time series in the $j$th block, resulting from integrating the local parameter $\theta_{i,j}^*$, i.e. 
\begin{equation}\label{eq:marg_dist}
	\mathcal M(\{y_{i,t} \mid t \in A_{i,j}\})  = \int \prod_{t = t_{i,j}^-}^{t_{i,j}^+}  \Law(y_{i,t}\mid y_{i,t-1}, \theta_{i,j}^*) P_0(\theta_{i,j}^*) \D\theta_{i,j}^*.
\end{equation}
Clearly, the tractability of~\eqref{eq:marg_dist} depends on specific choices of the data distribution $\Law(y\mid \cdots)$ and the prior for the local parameter $Q_0(\theta_{i,j}^*)$. In some common scenarios, the integral can be solved analytically, obtaining a closed form expression for the marginal distribution, as done for example in Web Appendix~\ref{subsec:application_time_series}. Alternatively, for more complex scenarios such integral can be approximated trough numerical solutions. See, e.g., Section~\ref{sec:epi}.


Finally, we can marginalise also the weights $\bm \pi$, as we are not interested on estimates of the mixing distribution but on the clustering, obtaining the following distribution
\begin{equation}\label{eq:mod_int}
	\begin{split}
		\Law(\mathcal Y, \mathcal R) = 
		\frac{\Gamma(\alpha^+)}{\Gamma(\alpha^+ + n)} \prod_{r = 1}^{2^{T-1}} \frac{\Gamma(\alpha_r + n_r)}{\Gamma(\alpha_r)} \prod_{\{i: \rho_i = \tilde\rho_r\}}\prod_{j=1}^{m_i}\mathcal M(\{y_{i,t} \mid t \in A_{i,j}\}),
	\end{split}
\end{equation}
where $\alpha^+ = \sum_{r=1}^{2^{T-1}} \alpha_r$, $\{i:\rho_i = \tilde \rho_r\}$ denotes the set of indices for which the latent order $\rho_i$ coincides with the $r$th element of $\tilde p$, and $n_r$ is the cardinality of $\{i:\rho_i = \tilde \rho_r\}$, i.e., number of observations associated with the cluster with latent order $\tilde\rho_r$. A model of the form in~\eqref{eq:mod_int} describes the convolution of a Dirichlet-categorical model with the marginal distributions of the data. We further remark that the partition of the data $\lambda$ is then given by the equivalence classes described by allocating observations to different latent orders, with $\lambda = \{B_1, \dots, B_k\}$ whereas $B_j = \{ i : \rho_i = \rho_{(j)}^\dagger \}$ is the $j$th block of the partitions whose latent orders are equal to $\rho_{(j)}^\dagger$, and $\{\rho_{(1)}^\dagger, \dots, \rho_{(k)}^\dagger\}$ denotes the set of orders out of $\tilde \rho_1, \dots, \tilde \rho_{2^{T-1}}$ with at least one observation assigned, i.e. the elements in $\{\rho_r : n_r > 0\}$. In the following, the model is elicited by considering and objective prior specification for the weights, i.e. $\alpha_r = \alpha$ for all $r = 1, \dots, 2^{T-1}$. We recall that the probability of having ties while sampling two observations from a Dirichlet-categorical model is given by 
\[
P(X_2 = X_1) = \frac{\alpha + 1}{2^{T-1} \alpha + 1}
\]
with $\lim_{\alpha \to 0} P(X_2 = X_1) = 1$ and $\lim_{\alpha \to +\infty} P(X_2 = X_1) = \frac{1}{2^{T-1}}$. Therefore, to ensure a significant probability of a tie when dealing with large values of $T$, in the following we set $\alpha = \frac{1}{2^{T-1}}$, in the spirit of \citet{Per47}. 

\section{Informed split-and-merge algorithm}\label{sec:algorithm}

The model we consider in the manuscript has a flexible structure which can accommodate for various types of data. The bottleneck of facing posterior inference is due to two main sources: the distribution we assume for the data $\Law(y_{i,t}\mid y_{i,t-1}, \theta_{i,t})$, which depends on specific analyses we are performing, and the cardinality of $\tilde p$. Regarding the latter, the number of latent orders scales poorly with the number of observed values for the time series, with $2^{T-1}$ possible orders when we observe $T$ distinct times. Even for a small $T$, we cannot explore the whole partitions' space in a reasonable time, and we need to use suitable computational strategies to obtain a sample from the target distribution of interest. 

Among the possible approaches, we build a sampling strategy upon the split-merge algorithm for nonparametric priors and clustering problems \citep{GreenPeterRichardson2001,JainNeal2004}, a particular type of MCMC algorithm that can be used to update the latent partition $\lambda$ of a model as in~\eqref{mod:spec}. Intuitively, at each step of the algorithm two indices of the latent partition, $i,\ell \in \{1, \dots, n\}$ with $i \neq \ell$, are randomly chosen. If the two indices belong to the same block of $\lambda$, meaning that the $i$th and the $\ell$th observations belong to the same cluster, their cluster is randomly splitted in two blocks. Otherwise, if the two indices $i$ and $\ell$ belong to different blocks of $\lambda$, the two clusters are merged. Once the split or the merge step is completed, we obtain a candidate partition $\lambda^{(N)}$ to update the current state of chain. We then perform a Metropolis--Hastings step to accept $\lambda^{(N)}$ or to stay on the previous state of the chain. Algorithm~\ref{algo:main_algorithm} shows the pseudo-code to produce a sample of size $M$ from the posterior distribution of the latent partition $\lambda$.

\begin{algorithm}[!ht]
	\setstretch{0.85}
	\DontPrintSemicolon
	\textbf{input}{ a partition $\lambda^{(0)} = \{B_1^{(0)}, \dots, B_k^{(0)}\}$ of $\{1, \dots, n\}$, initial values for the unique latent orders $\mathcal R^{*(0)}$.}
	
	\For{$m=1,\dots, M$}  { 
		
		\begin{enumerate}
			\item[a)] \textbf{set} $\lambda^{(N)} = \lambda^{(m-1)}$, with $\{B_1^{(N)}, \dots, B_k^{(N)}\}$ denoting the blocks of $\lambda^{(N)}$, \\ and $\mathcal R^{*(N)} = \mathcal R^{*(0)}$. 
			\item[b)] \textbf{sample} $i,\ell \in \{1,\dots,n\}$ such that $i \neq \ell$.
			\begin{itemize}
				\item[] \textbf{if} both $i$ and $\ell$ belong to the same block $B_s^{(m-1)}$, perform a split:
				\begin{enumerate}
					\item[i)] assign $i$ to $B_s^{(N)}$ and $\ell$ to a new block $B_{k+1}^{(N)}$;
					\item[ii)] assign randomly each values of $B_s^{(m-1)}$ to $B_s^{(N)}$ or $B_{k+1}^{(N)}$;
					\item[iii)] sample the two distinct unique values of the latent orders in $\mathcal R^{*(N)}$\\ associated with the observations whose indices belong to $B_s^{(N)}$ or $B_{k+1}^{(N)}$ from~\eqref{eq:proposal}.
				\end{enumerate}
				\item[] \textbf{else if} $i$ and $\ell$ belong to different blocks, with $i \in B_s^{(m-1)}$ and\\ $\ell \in B_w^{(m-1)}$, perform a merge:
				\begin{enumerate}
					\item[i)] assign all the indices in $B_w^{(m-1)}$ to $B_s^{(m-1)}$ and destroy $B_w^{(m-1)}$; 
					\item[ii)] sample the  unique value of the latent orders in $\mathcal R^{*(N)}$ associated \\ with all the observations whose indices are in $B_s^{(m-1)}$ from~\eqref{eq:proposal}.
				\end{enumerate}
			\end{itemize}
			
			\item[c)] \textbf{perform} a Metropolis--Hastings step to accept the proposed values. If accepted,\\ set $(\lambda^{(m)}, \mathcal R^{*(m)}) = (\lambda^{(N)}, \mathcal R^{*(N)})$. Set $(\lambda^{(m)}, \mathcal R^{*(m)}) = (\lambda^{(m-1)}, \mathcal R^{*(m-1)})$ otherwise.
			
			\item[d)] \textbf{perform} an acceleration step to update the unique values\\ $\rho_1^{*(m)}, \dots, \rho_k^{*(m)}$ in $\mathcal R^{*(m)}$.
			
		\end{enumerate}
		
	}
	\textbf{end}
	\caption{\label{algo:main_algorithm} Split-merge algorithm to sample from $\Law(\lambda \mid \mathcal Y)$}
\end{algorithm} 

The main challenge of using the split-merge algorithm in our framework is the sparse nature of $\tilde p$. When we perform a split or merge step, we need to propose a new latent order for each new block we create. Because $\tilde p$ has a large cardinality, randomly proposing those orders from a distribution independent of the data or previous state makes them unlikely to be representative of the new clusters, even when the proposed partition $\lambda^{(N)}$ is a suitable candidate.
To avoid this problem, we resort to an informed proposal for those orders. In particular, inspired by~\citet{Zha22}, we included in the proposal specification the information arising from the observed time series. Hence, our proposal equals 
\begin{equation} \label{eq:proposal}
	\psi(\rho \mid \mathcal Y) = \sum_{i=1}^n \frac{1}{n} \Law(\rho \mid \bm y_i),
\end{equation} 
i.e. a mixture of the posterior distribution of the latent orders conditionally on each observation separately. Here, $\Law(\rho\mid\bm y_i)$ denotes the posterior distribution of the latent order $\rho$ given the $i$th time series $\bm y_i$. 
Such mixture strategy for the proposal shrinks the probability
mass across the space of orders $\rho_1, \dots, \rho_{2^{T-1}}$ to the elements which are likely at least for a single observations, decreasing the probability of sampling values which are not representative of any observed time series. While sampling from a generic component of~\eqref{eq:proposal} is feasible, evaluating the entire mixture requires the computation of $n$ normalization constants, intractable for large $T$. These normalization constants can be computed before starting the algorithm. For example, we considered an importance sampling strategy with uniform importance distribution over the partitions' space (see Web Appendix~\ref{app:additional_fig_and_algo}). Further details on the implementation are deferred to the Web Appendix~\ref{appendix:main_algorithm}. Finally, Web Appendix~\ref{subsec:application_time_series} presents a simulation study with real-valued time series to validate Algorithm~\ref{algo:main_algorithm}, where subsets of synthetic data share the same structural change times and we assumed as kernel function for each block an univariate Ornstein-Uhlenbeck process \citep{Uhl30}. We can appreciate that posterior inference performed with Algorithm~\ref{algo:main_algorithm} provides reliable estimates of the latent partition of the data, and identify clusters which actually share the same change points.

\section{Clustering epidemics with similar structural changes} \label{sec:epi}


Next, we apply the previous method to epidemic data.  Hereby, we aim to model jointly epidemic spread in different areas, while clustering similar regions on the basis of structural changes in the spreading of the disease. 
To this end, we consider the standard compartmental susceptible-infected-removed (SIR) model, in which the population is segregated based on immunological statuses of the individuals. In an SIR model, individuals belong to exactly one of the three compartments of susceptible ($S$), infected and infectious ($I$) or recovered/removed ($R$) at any given time point. Under the stochastic law of mass-action \citep{AnderssonBritton2000}, an infected individual infects susceptible individuals whenever an individual Poisson clock rings, before eventually recovering. Once recovered, they play no role in the dynamics of the disease spread. 

Let $\bm X(t) = (X_S(t), X_I(t), X_R(t))$ where the stochastic processes  $X_S(t), X_I(t)$, and $ X_R(t)$ denote the numbers at time $t$ of susceptible, infected and recovered individuals respectively, and take values in the set of non-negative integers $\mathbb Z_+$. Since no birth or immigration is assumed into the population, the total population size is conserved at all times, i.e. $X_S(t)+ X_I(t)+ X_R(t) = X_S(0)+ X_I(0)+ X_R(0)$ at all times $t\ge 0$. Since we are primarily interested in large populations, we  use the quantity 
\begin{align}
	\varepsilon = \frac{1}{X_S(0)} \in (0, \infty)
\end{align}
as a scaling parameter and study the behaviour of the system as $\varepsilon \to 0$. We further assume $\varepsilon X_I(0) \to I_0 \in (0, 1) $ as $\varepsilon \to 0$, where the limiting quantity $I_0$ is non-random (deterministic), and assume the initial number of  recovered individuals $X_R(0)$ is zero. Let $\beta(t)$ and $\xi(t)$, bounded functions of time $t$,  represent the time-varying infection and recovery rates respectively.  Let us consider the scaled stochastic process $\bm X^{(\varepsilon)}(t) = (\eX_S(t), \eX_I(t), \eX_R(t))$, where $\eX_S = \varepsilon X_S, \eX_I = \varepsilon X_I$, and $\eX_R = \varepsilon X_R$. The scaled stochastic process $\bm X^{(\varepsilon)}(t)$ is a continuous-time Markov process satisfying 
\begin{align}
	\begin{aligned}
		\eX_S(t) &{}= \eX_S(0) - \eM_S(t) -   \int_0^t  \beta(u) \eX_S(u_-) \eX_I(u_-)  \D u, \\
		\eX_I(t) &{}= \eX_I(0) + \eM_I(t) + \int_0^t  \beta(u) \eX_S(u_-) \eX_I(u_-) \D u - \int_0^t \xi(u) \eX_I(u_-) \D u, \\
		\eX_R(t) &{}= \eX_R(0) + \eM_R(t) +   \int_0^t \xi(u) \eX_I(u_-)\D u,
	\end{aligned}
	\nonumber 
\end{align}
where $u_-$ denotes the left-hand limit at $u$ and the stochastic processes $\eM_S, \eM_I$, and $\eM_R$ are square-integrable zero-mean martingales, which converge to the zero function in probability as $\varepsilon \to 0$. See Web Appendix~\ref{app:epi} for a sketch of the derivation of these trajectory equations. 
Let $\norm{(x_1, x_2, x_3)}{\infty} = \max\{|{x_1}|, |x_2|, |x_3|\}$. As consequence of the functional law of large numbers, it can be proven that
\begin{align}
	\lim_{\varepsilon \to 0} \sup_{t \le T} \norm{\bm X^{(\varepsilon)}(t) - \bar{\bm X}(t) }{\infty} \stackrel{a.s.}{\longrightarrow} 0 
	\label{eq:SIR_limit}
\end{align}
where $\bar{\bm X}(t) = (S(t), I(t) , R(t))$ is the solution of the following system of ordinary differential equations
\begin{align}
	\begin{aligned}
		\frac{\mathrm{d}}{\mathrm{d}t}S = - \beta(t) S I, \qquad
		\frac{\mathrm{d}}{\mathrm{d}t}I = \beta(t) S I - \xi(t) I , \qquad
		\frac{\mathrm{d}}{\mathrm{d}t}R = \xi(t) I,
	\end{aligned}
	\label{eq:meanfield_SIR_ODE}
\end{align}
with the initial condition $S(0)=1, I(0) = I_0 $, and $R(0)=0$, and some $T>0$ \citep[see, e.g., Theorem 3.1 of][for a proof]{KhudaBukhsh2024SIR_SEIR}.

We note from~\eqref{eq:SIR_limit} that the cumulative infection pressure  $H_\varepsilon(t)$ converges to a deterministic limit, satisfying
\begin{align*}
	H_\varepsilon(t) = \int_0^t  \beta(u) \eX_I(u) \D u \stackrel{a.s.}{\longrightarrow} H(t) = \int_0^t \beta(u) I(u) \D u. 
\end{align*}
Therefore, by virtue of the Sellke construction \citep[e.g., see ][and also Web Appendix~\ref{app:epi}]{AnderssonBritton2000}, in the limit of the large population the probability that a randomly chosen susceptible individual is still susceptible at time $t$ is given by $\exp\left(-H(t)\right)$, which is precisely the function $S(t)$. That is, the function $S(t)$ can be interpreted as an improper survival function (improper since $\lim_{t\to\infty} S(t) >0$) describing the random variable time to infection of a randomly chosen susceptible individual. This forms the basis of the so called dynamical survival analysis \citep[DSA,][]{khudabukhsh2022projecting,Rempaa2023Handbook}. In practice, we can condition on a final observation time $T$ to get a proper survival function, which admits a density 
\begin{align}
	f_{T}(t) &{} = -\left(\frac{1 }{1- S(T)} \right) \frac{\mathrm{d}}{\mathrm{d}t}S(t) = \frac{\beta(t) S(t) I(t) }{1- S(T)}. 
	\label{eq:DSA_infection_time_density}
\end{align}

The density in~\eqref{eq:DSA_infection_time_density} describes the dynamics for a single population in continuous time. A discrete-time analog of the  dynamical stochastic analysis method has indeed been developed in \cite{Wascher2024IDSA} in the context of monitoring the spread of a disease in a closed population such as a student cohort on campus. Since we are interested in applying our method to data at the level of countries, we decide in favour of a simple model. In practice, this means we will discretize time and use the time-discretized model for clustering purposes at equally-spaced times $t \in\{ 1, \dots, T\}$, which in the following correspond to different days, whereas the infection rate parameter of the generic $i$th observed country $\bm \beta_i = \{\beta_{i,1}, \dots, \beta_{i,T}\}$ changes at certain times described by the specific latent order of the data, while the remaining parameters are assumed to be homogeneous over time. Therefore, given a random sample of daily cases of infections $y_{i,1}, \dots, y_{i,T}$ at a generic $i$th location, observed up to some terminal time $T>0$, the likelihood contribution of the $i$th observed series is given by
\begin{align} \label{eq:lkl_epi}
	\Law(\bm y_i \mid \bm \beta_i, \rho_i) &= \prod_{j=1}^{m_i}\prod_{t=t_{i,j}^-}^{t_{i,j}^+} f_{i,T}(t)^{y_{i,t}},\quad i=1, \dots, n, 
\end{align} 
where $f_{i,T}(t)$ is defined in \eqref{eq:DSA_infection_time_density} and obtained by integrating numerically the system in~\eqref{eq:meanfield_SIR_ODE} with infection rate $\bm \beta_i$. This is the likelihood function we will use for our clustering inference.

We remark that, also for the epidemiological case, we are not interested in inferring parameter values but on the latent clustering structure of the data. However, while in Web Appendix~\ref{subsec:application_time_series} we could integrate analytically the parameters of the kernel function, here we resort to numerical integration of those parameters. Specifically, we assume the recovery rate $\xi$ to be constant and shared across different infection time series. The parameter $I_0$ can be updated through an observation-specific Metropolis--Hastings step. Finally, the infection rates $\bm\beta_i$s, which change over time, are marginalised via Monte Carlo integration. 


\subsection{Synthetic infection study} \label{subsec:sim_epi}

\begin{figure}[h]
	\centering
	\includegraphics[scale = 0.53]{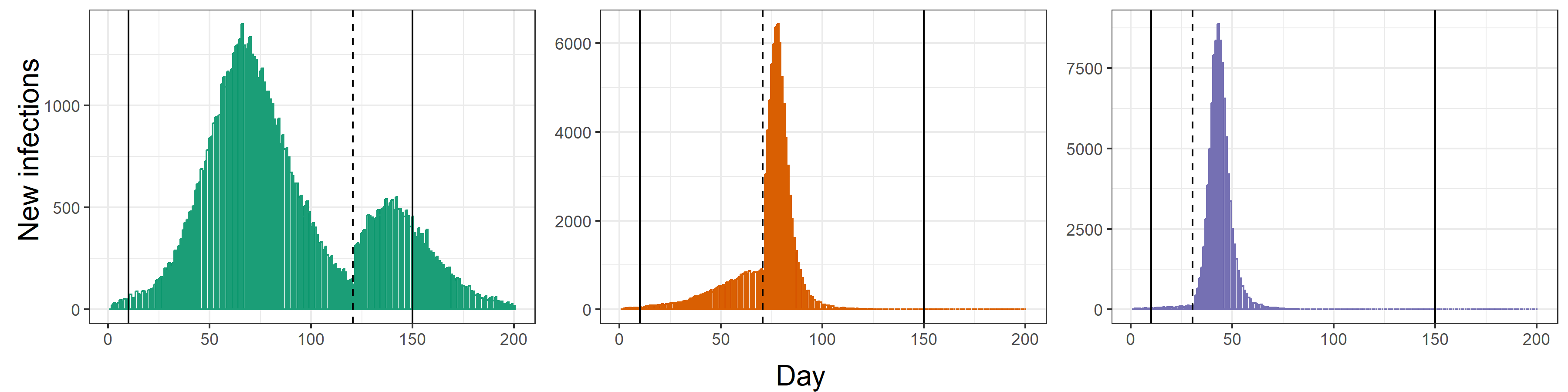}
	\caption{
		An example of bar plots of daily new cases in the synthetic study. Different panels correspond to different populations, one for each cluster considered in the study.  
	}
	\label{fig:sim_EPI_histograms}
\end{figure}

We first show through a simulation study the performance of the proposed model in an epidemiological scenario. Data are generated using the Doob--Gillespie algorithm \citep[see][and Web Appendix~\ref{app:additional_fig_and_algo} for pseudo-code]{Anderson2015StochasticAnalysis}. We consider a set of $n = 10$ populations, where the starting number of susceptibles is $S_0 = 100\,000$. We keep the recovery rate fixed at $\xi = 1/8$, we consider a time-varying rate of infection $\beta_{i,t}$ and a different starting proportion of infected $I_{i,0}$ for each population. The simulated data are generated in a time period of $200$ days, however in the analysis we consider a restricted window of time to remove the tails of the epidemic. We restrict the window from time $t = 10$ to time $t = 150$ thus resulting in a time period of length $T = 140$. We assume three different groups where the parameters are as shown in Table \ref{tab:sim_EPI_parameters} of Web Appendix~\ref{app:additional_fig_and_algo} . In the first group we simulate an epidemic where we have two large spreads, while in the second and third group we simulate a single large spread but with different change points. Figure \ref{fig:sim_EPI_histograms} shows an example of simulated data for each of the groups considered in the study.

Similarly to the time series synthetic study of Web Appendix~\ref{subsec:application_time_series}, we investigate how different parameters affect posterior estimates and the algorithm performance. Specifically, we vary the number of points used to marginalize via Monte Carlo the observation-specific infection rates, here denoted by MC, as well as the accuracy of the normalization constant $\mathrm B$ and depth of the proposal $\mathrm L$. We set $\xi = 1/8$ to be consistent with the parameters we used to generate synthetic data. Further, we marginalize the infection rate whereas the Monte Carlo integration is performed by sampling the unique values in $\bm \beta_i$ from a gamma distribution with $(4,10)$ as shape and rate parameters. 
Finally, for each population we randomly sample without replacement 20\% of the infection times, since the dynamical survival analysis method requires only a subset of infections \citep[see][]{khudabukhsh2022projecting}.
In each replicate, we run the algorithm for $5\,000$ iterations, of which $2\,000$ were discarded as burn-in period.  The point estimate of the latent partition $\hat\lambda$ is then obtained as minimizer of the Binder loss function \citep{Binder1978} among those visited by the algorithm. To assess the algorithm performance and the quality of the posterior estimates the Binder loss (BI) between the point estimate and the true latent partition of the data ($\mathrm{BI}$), which are reported in Table~\ref{tab:simulations_results_EPI}.

\begin{table}[h]
	\centering
	\begin{tabular}{cccccccccccc}
		\toprule
		\multicolumn{3}{c}{$\mathrm{MC} = 250$} & \multicolumn{3}{c}{$\mathrm{MC} = 500$}  & \multicolumn{3}{c}{$\mathrm{MC} = 1\,000$}  \\
		\cmidrule(lr){1-3}\cmidrule(lr){4-6} \cmidrule(lr){7-9} 
		$\mathrm B$ & $\mathrm L$ & $\mathrm{BI}(\hat \lambda, \lambda_0)$  & $\mathrm B$ & $\mathrm L$ & $\mathrm{BI}(\hat \lambda, \lambda_0)$  & $\mathrm B$ & $\mathrm L$ & $\mathrm{BI}(\hat \lambda, \lambda_0)$ \\ \midrule
		\multirow{3}{*}{1\,000}  
		& 1  & 0.345 & \multirow{3}{*}{1\,000} & 1 & 0.200 & \multirow{3}{*}{1\,000} & 1 & 0.005 \\
		& 25  & 0.294 & & 25  & 0.144 & & 25 & 0.010 \\
		& 100 & 0.257 & & 100 & 0.194 & & 100 & 0.018 \\ 
		\midrule
		\multirow{3}{*}{10\,000}  
		& 1 & 0.354 & \multirow{3}{*}{10\,000} & 1 & 0.234  & \multirow{3}{*}{10\,000} & 1 & $\ll10^{-4}$ \\
		& 25 & 0.323 & & 25  & 0.042 & & 25 & 0.010 \\
		& 100 & 0.281 & & 100 & 0.162 & & 100 & 0.008 \\
		\midrule
		\multirow{3}{*}{100\,000}  
		& 1  & 0.320 & \multirow{3}{*}{100\,000} & 1 & 0.280 & \multirow{3}{*}{100\,000} & 1 & 0.014 \\
		& 25  & 0.310 &  & 25  & 0.249 &  & 25 & 0.007 \\
		& 100 & 0.212 & & 100 & 0.092 &  & 100 & 0.009 \\
		\bottomrule
	\end{tabular}
	\caption{Posterior summaries of the synthetic epidemiological data. We consider different scenarios, where $\mathrm{MC}$ denotes the accuracy of the numerical integration of the infection rates, $\mathrm B$ denotes the accuracy of the normalization constant and $\mathrm L$ the proposal depth. The table reports the Binder loss between the point estimate and the true latent partition of the data (BI). Results are averaged over 50 replicates. 
	}
	\label{tab:simulations_results_EPI}
\end{table}
We can appreciate a similar behaviour to the one observed in the time series synthetic study: also with a complex kernel function, such as an SIR model, posterior estimates and performance of the algorithm are not particularly affected by the normalization constant accuracy $\mathrm B$ and the proposal depth $\mathrm L$. However, increasing the number of points used for the Monte Carlo marginalization of the infection rate (MC) has a positive effect on the quality of the partition estimates. In particular, for $\mathrm MC = 1000$ we obtain posterior estimates having comparable precision to the time series synthetic study (Web Appendix~\ref{subsec:application_time_series}), where local parameters are marginalized analytically. 


\subsection{SARS-CoV-2 Europe infection data}
\label{sec:applications_real_data}

We now consider the real data analysis motivating the model we study in this manuscript. Our aim is to identify possible clusters  of homogeneous states among the 27 countries that comprise the European Union, with respect to the spread of SARS-CoV-2 virus. The first confirmed case was in Bordeaux (France) on the 24th January 2020, while the first major outbreak was experienced in the late February by Italy, and different countries within the European Union experienced different starting time of the spreading. Even if the response to the pandemic was different from one country to another, some actions were taken jointly with the European commission such as travel ban and vaccination. Hereby, we consider infection data taken from the \cite{dataCovid19} database on COVID-19 daily new cases, for a time period of more than one year that goes from 1st June 2021 to 30st September 2022. We selected this specific time windows to mitigate differences of the pandemic, as it starts after the beginning of the world mass immunization campaigns and includes the largest spread of new detected case in Europe that lasted from the end of 2021 to September 2022.
We smooth the data by computing for each day the 7-days rolling average of the number of infected. By doing so, we remove some common measurement errors that occur when counting the number of daily new cases, e.g. the number of weekend recorded daily new cases is always smaller than during week days. 
We set $\xi = 1/6$ and the unique values of the infection rates out of each $\bm\beta_i$ are sampled from a gamma distribution with with $(2,10)$ as shape and rate parameters. 
We increase the accuracy of the normalization constants to $B = 100\,000$, since we have $T = 487$ days, we set $L=1$ to reduce the computational time, and $MC = 1\,000$. Finally, we do not consider the entire data, but a subset of $50\,000$ infection times randomly sampled for each country. 
The MCMC algorithm is run for $12\,500$ iterations of which $2\,500$ have been discarded as burn-in period. 
\begin{figure}[h] 
	\footnotesize
	\centering
	\includegraphics[width = 0.49\textwidth]{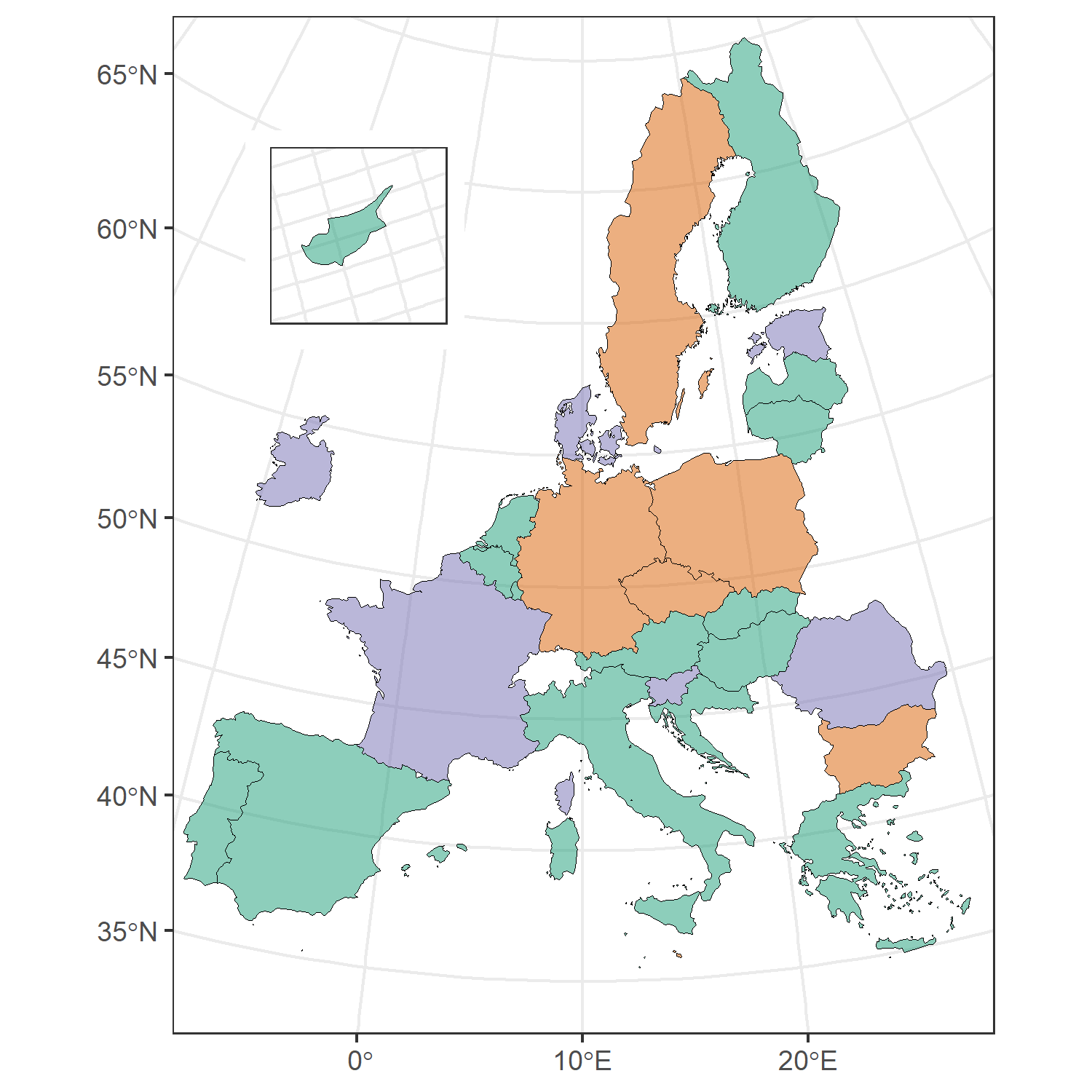}
	\includegraphics[width = 0.49\textwidth]{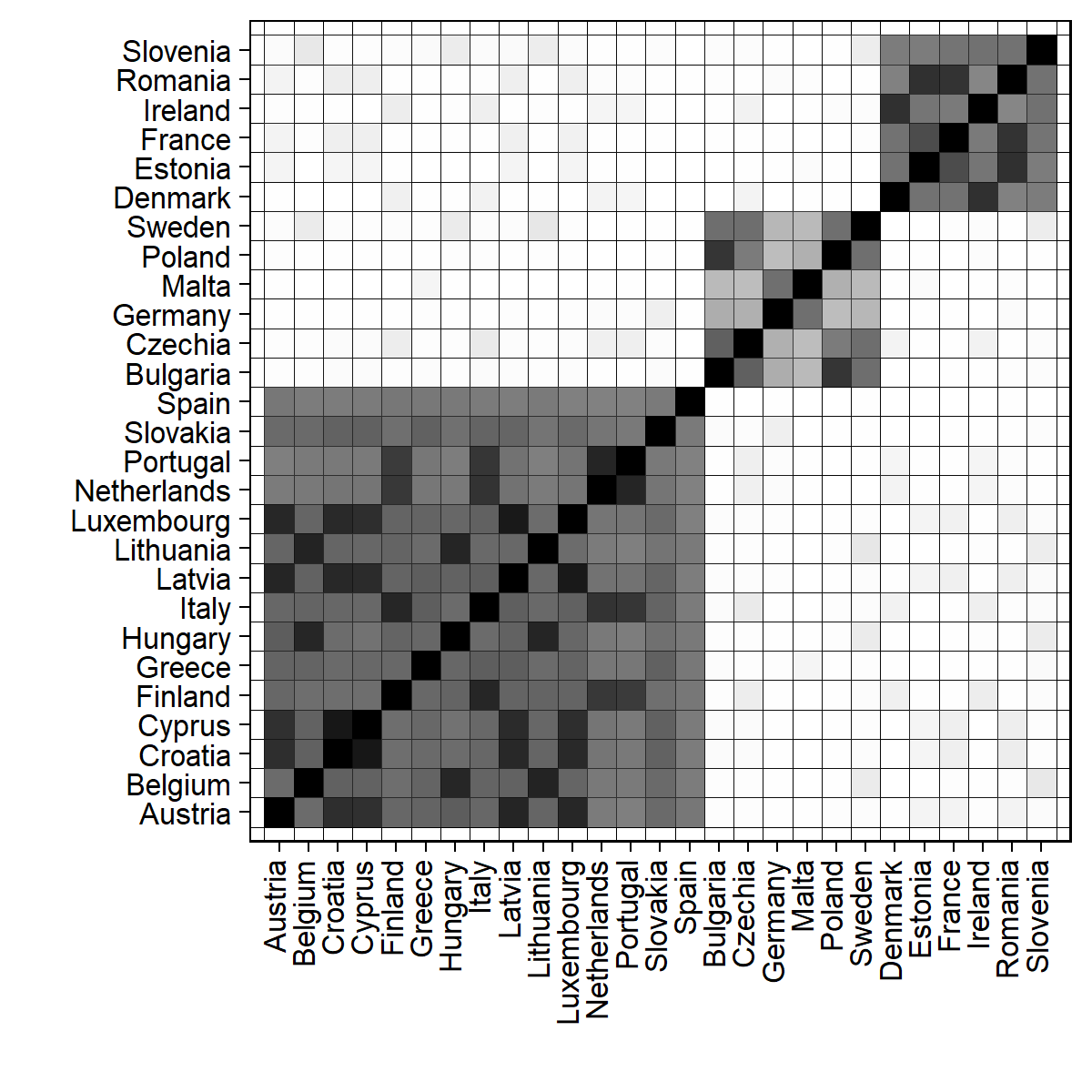}
	
	\caption{Summaries of the EU27 countries analysis. Left plot: countries colored consistently with the point estimate of the latent partition. Right plot: posterior similarity matrix. }
	\label{fig:COVID_real_data}
\end{figure}
The posterior point estimates of the latent partition in the data is composed by three clusters, which are shown Figure \ref{fig:COVID_real_data} along with the posterior similarity matrix. We have a larger cluster with 15 countries and two smaller clusters both with 6 countries. We can notice some geographical correlations in the green and orange clusters. For example, all the Mediterranean countries except of France and Slovenia belong to the larger cluster, while Germany, Poland and Czechia to the second cluster. Countries in the third cluster are more scattered and do not show any spatial relationships. The posterior similarity matrix in Figure \ref{fig:COVID_real_data} shows that the posterior distribution of the partition is centered around the selected partition. All the clusters are well defined, only the central cluster in the posterior similarity matrix shows lower similarities with respect to the other groups.

\section{Discussion} \label{sec:discussions}

In the previous sections, motivated by an epidemiological study, we introduced a novel approach to cluster together time-dependent observations having the same structural change times. Such approach works on a mixture representation of the model, whereas the mixing parameters is a random order of the observational times, which induce locally homogeneous behaviours but also change points. Beyond the cases we considered in the manuscript, this modeling approach can be easily adapted to several frameworks where the interest lies in defining groups of time-dependent observations. 

Along the manuscript, we have assumed a symmetric prior distribution. However, this assumption can be relaxed by considering prior information associated with specific latent orders, in the spirit of~\cite{Can17}. Further, instead of a Dirichlet-multinomial model, one can consider approaches based on more flexible finite-dimensional discrete prior distributions, in the spirit of~\cite{Lij20,Lij24}. However, the computational complexity of these approaches may result in models for time series clustering that are challenging to handle.

Further, there is no borrowing of information across different cluster.
While this assumption improves the model's tractability, having possibly dependent orders increase its flexibility. For example, recent contributions in this direction can be found in~\cite{Quinlan-2022}. 
In this manuscript we considered an approach based on the inspiring studies of~\cite{Mar14}, where the authors restrict an EPPF from the partitions' space to the orders' space. In the same spirit, one can consider to restrict a partially exchangeable partition probability function \citep[see, e.g.,][]{Pit95} from the partitions' space to the orders' space, obtaining a distribution which introduce dependence among different atoms. The model discussed in this paper can be then extended by considering such a distribution to introduce dependencies among different dimensions, but preserving the same combinatorial structure and flexibility.

\section*{Acknowledgements}

The first, second and fourth authors gratefully acknowledge the financial support from the Italian Ministry of University and Research (MUR), “Dipartimenti di Eccellenza” grant 2023-2027, and the DEMS Data Science Lab for supporting this work through computational resources. Wasiur R. KhudaBukhsh was partially supported by the Engineering and Physical Sciences Research Council (EPSRC), UK Research and Innovation (UKRI), grant number EP/Y027795/1.\vspace*{-8pt}



\bibliographystyle{apa} 
\bibliography{references.bib}

\begin{thebibliography}{}

\bibitem[\protect\astroncite{Aghabozorgi et~al.}{2015}]{AGHABOZORGI201516}
Aghabozorgi, S., {Seyed Shirkhorshidi}, A., and {Ying Wah}, T. (2015).
\newblock Time-series clustering – a decade review.
\newblock {\em Information Systems}, 53:16--38.

\bibitem[\protect\astroncite{Anderson and
  Kurtz}{2015}]{Anderson2015StochasticAnalysis}
Anderson, D.~F. and Kurtz, T.~G. (2015).
\newblock {\em Stochastic Analysis of Biochemical Systems}.
\newblock Springer International Publishing.

\bibitem[\protect\astroncite{Andersson and
  Britton}{2000}]{AnderssonBritton2000}
Andersson, H. and Britton, T. (2000).
\newblock {\em Stochastic Epidemic Models and Their Statistical Analysis}.
\newblock Springer New York.

\bibitem[\protect\astroncite{Applebaum}{2019}]{Applebaum2019Semigroups}
Applebaum, D. (2019).
\newblock {\em Semigroups of Linear Operators: With Applications to Analysis,
  Probability and Physics}.
\newblock Cambridge University Press.

\bibitem[\protect\astroncite{Baker et~al.}{2022}]{Baker2022Analyzing}
Baker, E., Barbillon, P., Fadikar, A., Gramacy, R.~B., Herbei, R., Higdon, D.,
  Huang, J., Johnson, L.~R., Ma, P., Mondal, A., Pires, B., Sacks, J., and
  Sokolov, V. (2022).
\newblock {Analyzing Stochastic Computer Models: A Review with Opportunities}.
\newblock {\em Statistical Science}, 37(1):64 -- 89.

\bibitem[\protect\astroncite{Barry and Hartigan}{1992}]{Bar92}
Barry, D. and Hartigan, J.~A. (1992).
\newblock {Product Partition Models for Change Point Problems}.
\newblock {\em The Annals of Statistics}, 20(1):260 -- 279.

\bibitem[\protect\astroncite{Barry and Hartigan}{1993}]{Bar93}
Barry, D. and Hartigan, J.~A. (1993).
\newblock A bayesian analysis for change point problems.
\newblock {\em Journal of the American Statistical Association},
  88(421):309--319.

\bibitem[\protect\astroncite{Becker}{1993}]{Becker1993Parametric}
Becker, N.~G. (1993).
\newblock Parametric inference for epidemic models.
\newblock {\em Mathematical Biosciences}, 117(1):239--251.

\bibitem[\protect\astroncite{Bhatt et~al.}{2023}]{Bhatt2023Semimechanistic}
Bhatt, S., Ferguson, N., Flaxman, S., Gandy, A., Mishra, S., and Scott, J.~A.
  (2023).
\newblock Semi-mechanistic bayesian modelling of covid-19 with renewal
  processes.
\newblock {\em Journal of the Royal Statistical Society Series A: Statistics in
  Society}, 186(4):601–615.

\bibitem[\protect\astroncite{Binder}{1978}]{Binder1978}
Binder, D.~A. (1978).
\newblock Bayesian cluster analysis.
\newblock {\em Biometrika}, 65(1):31--38.

\bibitem[\protect\astroncite{Bong et~al.}{2023}]{Bong2023Frequentist}
Bong, H., Ventura, V., and Wasserman, L. (2023).
\newblock Frequentist inference for semi-mechanistic epidemic models with
  interventions.

\bibitem[\protect\astroncite{Britton and Pardoux}{2019}]{Britton2019Stochastic}
Britton, T. and Pardoux, E. (2019).
\newblock {\em {Stochastic Epidemic Models with Inference}}.
\newblock Springer Cham.

\bibitem[\protect\astroncite{Brodsky and
  Darkhovsky}{1993}]{brodsky1993nonparametric}
Brodsky, E. and Darkhovsky, B. (1993).
\newblock {\em Nonparametric Methods in Change Point Problems}.
\newblock Springer Netherlands.

\bibitem[\protect\astroncite{Canale et~al.}{2017}]{Can17}
Canale, A., Lijoi, A., Nipoti, B., and Prünster, I. (2017).
\newblock {On the Pitman–Yor process with spike and slab base measure}.
\newblock {\em Biometrika}, 104(3):681--697.

\bibitem[\protect\astroncite{{Change Points Clustering}}{2024}]{lucarepository}
{Change Points Clustering} (2024).
\newblock {Repository with R and C++ code to perform clustering of
  time-dependent observations with common changes in time.}
\newblock \url{https://github.com/lucadanese/Change_Points_Clustering}.

\bibitem[\protect\astroncite{Chen and Gupta}{2012}]{ChenJieGuptaArjun2021}
Chen, J. and Gupta, A. (2012).
\newblock {\em Parametric Statistical Change Point Analysis}.
\newblock Birkhäuser Boston.

\bibitem[\protect\astroncite{Chen and Gupta}{1997}]{ChenGupta1997}
Chen, J. and Gupta, A.~K. (1997).
\newblock Testing and locating variance changepoints with application to stock
  prices.
\newblock {\em Journal of the American Statistical Association},
  92(438):739--747.

\bibitem[\protect\astroncite{Chernoff and Zacks}{1964}]{Cher64}
Chernoff, H. and Zacks, S. (1964).
\newblock {Estimating the Current Mean of a Normal Distribution which is
  Subjected to Changes in Time}.
\newblock {\em The Annals of Mathematical Statistics}, 35(3):999 -- 1018.

\bibitem[\protect\astroncite{Chitwood
  et~al.}{2022}]{Chitwood2022Reconstructing}
Chitwood, M.~H., Russi, M., Gunasekera, K., Havumaki, J., Klaassen, F., Pitzer,
  V.~E., Salomon, J.~A., Swartwood, N.~A., Warren, J.~L., Weinberger, D.~M.,
  Cohen, T., and Menzies, N.~A. (2022).
\newblock Reconstructing the course of the covid-19 epidemic over 2020 for us
  states and counties: Results of a bayesian evidence synthesis model.
\newblock {\em PLOS Computational Biology}, 18(8):e1010465.

\bibitem[\protect\astroncite{Choi and Rempala}{2011}]{Choi2011Inference}
Choi, B. and Rempala, G.~A. (2011).
\newblock {Inference for discretely observed stochastic kinetic networks with
  applications to epidemic modeling}.
\newblock {\em Biostatistics}, 13(1):153--165.

\bibitem[\protect\astroncite{Corradin et~al.}{2022}]{corradin2022}
Corradin, R., Danese, L., and Ongaro, A. (2022).
\newblock Bayesian nonparametric change point detection for multivariate time
  series with missing observations.
\newblock {\em International Journal of Approximate Reasoning}, 143:26--43.

\bibitem[\protect\astroncite{Di~Lauro et~al.}{2022}]{DiLauro2022NonMarkov}
Di~Lauro, F., KhudaBukhsh, W.~R., Kiss, I.~Z., Kenah, E., Jensen, M., and
  Rempa{\l}a, G.~A. (2022).
\newblock Dynamic survival analysis for non-markovian epidemic models.
\newblock {\em Journal of The Royal Society Interface}, 19(191):20220124.

\bibitem[\protect\astroncite{Engel and
  Nagel}{2000}]{EngelNagel2000OneParameter}
Engel, K.-J. and Nagel, R. (2000).
\newblock {\em One-Parameter Semigroups for Linear Evolution Equations}.
\newblock Springer-Verlag.

\bibitem[\protect\astroncite{{European Centre for Disease Prevention and
  Controls}}{2022}]{dataCovid19}
{European Centre for Disease Prevention and Controls} (2022).
\newblock {Database on COVID-19 daily new cases}.
\newblock
  \url{https://www.ecdc.europa.eu/en/publications-data/data-daily-new-cases-covid-19-eueea-country}.
  Dataset from October 2022 is no longer updated.

\bibitem[\protect\astroncite{Fadikar et~al.}{2018}]{Fadikar2018Calibrating}
Fadikar, A., Higdon, D., Chen, J., Lewis, B., Venkatramanan, S., and Marathe,
  M. (2018).
\newblock Calibrating a stochastic, agent-based model using quantile-based
  emulation.
\newblock {\em SIAM/ASA Journal on Uncertainty Quantification},
  6(4):1685–1706.

\bibitem[\protect\astroncite{Fr{\"u}hwirth-Schnatter and
  Pamminger}{2010}]{Frühwirth-Schnatter-Pamminger-2008}
Fr{\"u}hwirth-Schnatter, S. and Pamminger, C. (2010).
\newblock {Model-based clustering of categorical time series}.
\newblock {\em Bayesian Analysis}, 5(2):345 -- 368.

\bibitem[\protect\astroncite{Frühwirth-Schnatter and
  Kaufmann}{2008}]{Frühwirth-Schnatter-2008}
Frühwirth-Schnatter, S. and Kaufmann, S. (2008).
\newblock Model-based clustering of multiple time series.
\newblock {\em Journal of Business \& Economic Statistics}, 26(1):78--89.

\bibitem[\protect\astroncite{Fuentes–García et~al.}{2010}]{Fue10}
Fuentes–García, R., Mena, R., and Walker, S. (2010).
\newblock {A Probability for Classification Based on the Dirichlet Process
  Mixture Model}.
\newblock {\em Journal of Classification}, 27:389--403.

\bibitem[\protect\astroncite{Giampino
  et~al.}{2024}]{giampino2024localleveldynamicrandom}
Giampino, A., Guindani, M., Nipoti, B., and Vannucci, M. (2024).
\newblock Local level dynamic random partition models for changepoint
  detection.

\bibitem[\protect\astroncite{Gikhman and
  Skorokhod}{2004}]{Gikhman2004StochPro2}
Gikhman, I.~I. and Skorokhod, A.~V. (2004).
\newblock {\em The Theory of Stochastic Processes II}.
\newblock Springer Berlin Heidelberg.

\bibitem[\protect\astroncite{Gleeson et~al.}{2021}]{Gleeson2021Calibrating}
Gleeson, J.~P., Brendan~Murphy, T., O’Brien, J.~D., Friel, N., Bargary, N.,
  and O’Sullivan, D. J.~P. (2021).
\newblock Calibrating covid-19 susceptible-exposed-infected-removed models with
  time-varying effective contact rates.
\newblock {\em Philosophical Transactions of the Royal Society A: Mathematical,
  Physical and Engineering Sciences}, 380(2214).

\bibitem[\protect\astroncite{Green and
  Richardson}{2001}]{GreenPeterRichardson2001}
Green, P.~J. and Richardson, S. (2001).
\newblock Modelling heterogeneity with and without the dirichlet process.
\newblock {\em Scandinavian Journal of Statistics}, 28(2):355--375.

\bibitem[\protect\astroncite{Hartigan}{1990}]{Har90}
Hartigan, J. (1990).
\newblock Partition models.
\newblock {\em Communications in Statistics - Theory and Methods},
  19(8):2745--2756.

\bibitem[\protect\astroncite{Hong and Li}{2020}]{Hong2020Estimation}
Hong, H.~G. and Li, Y. (2020).
\newblock Estimation of time-varying reproduction numbers underlying
  epidemiological processes: A new statistical tool for the covid-19 pandemic.
\newblock {\em PLOS ONE}, 15(7):e0236464.

\bibitem[\protect\astroncite{Huang et~al.}{2024}]{Hua24}
Huang, J., Morsomme, R., Dunson, D., and Xu, J. (2024).
\newblock Detecting changes in the transmission rate of a stochastic epidemic
  model.
\newblock {\em Statistics in Medicine}, 43(10):1867--1882.

\bibitem[\protect\astroncite{Inclan and Tiao}{1994}]{InclanTiao1994}
Inclan, C. and Tiao, G.~C. (1994).
\newblock Use of cumulative sums of squares for retrospective detection of
  changes of variance.
\newblock {\em Journal of the American Statistical Association},
  89(427):913--923.

\bibitem[\protect\astroncite{Jain and Neal}{2004}]{JainNeal2004}
Jain, S. and Neal, R.~M. (2004).
\newblock A split-merge markov chain monte carlo procedure for the dirichlet
  process mixture model.
\newblock {\em Journal of Computational and Graphical Statistics},
  13(1):158--182.

\bibitem[\protect\astroncite{Jain and Neal}{2007}]{JainNeal2007}
Jain, S. and Neal, R.~M. (2007).
\newblock {Splitting and merging components of a nonconjugate Dirichlet process
  mixture model}.
\newblock {\em Bayesian Analysis}, 2(3):445 -- 472.

\bibitem[\protect\astroncite{Juárez and Steel}{2010}]{Juarez-Steel-2010}
Juárez, M.~A. and Steel, M. F.~J. (2010).
\newblock Model-based clustering of non-gaussian panel data based on skew- t
  distributions.
\newblock {\em Journal of Business \& Economic Statistics}, 28(1):52--66.

\bibitem[\protect\astroncite{Kermack and McKendrick}{1927}]{Kermack1927}
Kermack, W.~O. and McKendrick, A.~G. (1927).
\newblock A contribution to the mathematical theory of epidemics.
\newblock {\em Proceedings of the Royal Society of London. Series A, Containing
  Papers of a Mathematical and Physical Character}, 115(772):700–721.

\bibitem[\protect\astroncite{KhudaBukhsh
  et~al.}{2023}]{khudabukhsh2022projecting}
KhudaBukhsh, W.~R., Bastian, C.~D., Wascher, M., Klaus, C., Sahai, S.~Y., Weir,
  M.~H., Kenah, E., Root, E., Tien, J.~H., and Rempa{\l}a, G.~A. (2023).
\newblock Projecting covid-19 cases and hospital burden in ohio.
\newblock {\em Journal of Theoretical Biology}, 561:111404.

\bibitem[\protect\astroncite{KhudaBukhsh et~al.}{2020}]{KhudaBukhsh2020DSA}
KhudaBukhsh, W.~R., Choi, B., Kenah, E., and Rempała, G.~A. (2020).
\newblock Survival dynamical systems: individual-level survival analysis from
  population-level epidemic models.
\newblock {\em Interface Focus}, 10(1):20190048.

\bibitem[\protect\astroncite{KhudaBukhsh and
  Rempała}{2024}]{KhudaBukhsh2024SIR_SEIR}
KhudaBukhsh, W.~R. and Rempała, G.~A. (2024).
\newblock How to correctly fit an sir model to data from an seir model?
\newblock {\em Mathematical Biosciences}, 375:109265.

\bibitem[\protect\astroncite{Kypraios et~al.}{2017}]{Kypraios2017tutorial}
Kypraios, T., Neal, P., and Prangle, D. (2017).
\newblock A tutorial introduction to bayesian inference for stochastic epidemic
  models using approximate bayesian computation.
\newblock {\em Mathematical Biosciences}, 287:42--53.

\bibitem[\protect\astroncite{Lijoi et~al.}{2020}]{Lij20}
Lijoi, A., Prünster, I., and Rigon, T. (2020).
\newblock {The Pitman–Yor multinomial process for mixture modelling}.
\newblock {\em Biometrika}, 107(4):891--906.

\bibitem[\protect\astroncite{Lijoi et~al.}{2024}]{Lij24}
Lijoi, A., Prünster, I., and Rigon, T. (2024).
\newblock Finite-dimensional discrete random structures and bayesian
  clustering.
\newblock {\em Journal of the American Statistical Association},
  119(546):929--941.

\bibitem[\protect\astroncite{Loschi and Cruz}{2002}]{Los02}
Loschi, R. and Cruz, F. (2002).
\newblock An analysis of the influence of some prior specifications in the
  identification of change points via product partition model.
\newblock {\em Computational Statistics \& Data Analysis}, 39(4):477--501.

\bibitem[\protect\astroncite{Martínez and Mena}{2014}]{Mar14}
Martínez, A.~F. and Mena, R.~H. (2014).
\newblock {On a Nonparametric Change Point Detection Model in Markovian
  Regimes}.
\newblock {\em Bayesian Analysis}, 9(4):823 -- 858.

\bibitem[\protect\astroncite{Niels~G.}{1993}]{Becker1993Martingale}
Niels~G., B. (1993).
\newblock Martingale methods for the analysis of epidemic data.
\newblock {\em Statistical Methods in Medical Research}, 2(1):93--112.

\bibitem[\protect\astroncite{Niu et~al.}{2016}]{NiuHangZhang2016}
Niu, Y.~S., Hao, N., and Zhang, H. (2016).
\newblock Multiple change-point detection: A selective overview.
\newblock {\em Statistical Science}, 31(4):611--623.

\bibitem[\protect\astroncite{Page}{1954}]{Pag54}
Page, E.~S. (1954).
\newblock {Continuous inspection schemes}.
\newblock {\em Biometrika}, 41(1-2):100--115.

\bibitem[\protect\astroncite{Page}{1957}]{Pag57}
Page, E.~S. (1957).
\newblock {On problems in which a change in a parameter occurs at an unknown
  point}.
\newblock {\em Biometrika}, 44(1-2):248--252.

\bibitem[\protect\astroncite{Page et~al.}{2022}]{Page2022}
Page, G.~L., Quintana, F.~A., and Dahl, D.~B. (2022).
\newblock Dependent modeling of temporal sequences of random partitions.
\newblock {\em Journal of Computational and Graphical Statistics},
  31(2):614--627.

\bibitem[\protect\astroncite{Pedroso et~al.}{2023}]{Pedroso-2021}
Pedroso, R.~C., Loschi, R.~H., and Quintana, F.~A. (2023).
\newblock {Multipartition model for multiple change point identification}.
\newblock {\em TEST}, 32(2):759--783.

\bibitem[\protect\astroncite{Perks}{1947}]{Per47}
Perks, W. (1947).
\newblock Some observations on inverse probability including a new indifference
  rule.
\newblock {\em Journal of the Institute of Actuaries}, 73(2):285--334.

\bibitem[\protect\astroncite{Pitman}{1995}]{Pit95}
Pitman, J. (1995).
\newblock Exchangeable and partially exchangeable random partitions.
\newblock {\em Probability Theory and Related Fields}, 102(2):145--158.

\bibitem[\protect\astroncite{Pitman}{2006}]{Pit06}
Pitman, J. (2006).
\newblock {\em Combinatorial stochastic processes}.
\newblock Springer Berlin.

\bibitem[\protect\astroncite{Quinlan et~al.}{2024}]{Quinlan-2022}
Quinlan, J.~J., Page, G.~L., and Castro, L.~M. (2024).
\newblock {Joint Random Partition Models for Multivariate Change Point
  Analysis}.
\newblock {\em Bayesian Analysis}, 19(1):21 -- 48.

\bibitem[\protect\astroncite{Quintana and Iglesias}{2003}]{Qui03}
Quintana, F.~A. and Iglesias, P.~L. (2003).
\newblock Bayesian clustering and product partition models.
\newblock {\em Journal of the Royal Statistical Society: Series B (Statistical
  Methodology)}, 65(2):557--574.

\bibitem[\protect\astroncite{Raúl~Fierro and
  Balakrishnan}{2015}]{Fierro2015Statistical}
Raúl~Fierro, V.~L. and Balakrishnan, N. (2015).
\newblock Statistical inference on a stochastic epidemic model.
\newblock {\em Communications in Statistics - Simulation and Computation},
  44(9):2297--2314.

\bibitem[\protect\astroncite{Rempala and
  KhudaBukhsh}{2023}]{Rempaa2023Handbook}
Rempala, G.~A. and KhudaBukhsh, W.~R. (2023).
\newblock {\em Dynamical Survival Analysis for Epidemic Modeling}, page 1–17.
\newblock Springer International.

\bibitem[\protect\astroncite{Same et~al.}{2011}]{same-2011}
Same, A., Chamroukhi, F., Govaert, G., and Aknin, P. (2011).
\newblock Model-based clustering and segmentation of time series with change in
  regime.
\newblock {\em Advances in Data Analysis and Classification}, 5:301--321.

\bibitem[\protect\astroncite{Seymour et~al.}{2022}]{Seymour2022Bayesian}
Seymour, R.~G., Kypraios, T., and O’Neill, P.~D. (2022).
\newblock Bayesian nonparametric inference for heterogeneously mixing
  infectious disease models.
\newblock {\em Proceedings of the National Academy of Sciences},
  119(10):e2118425119.

\bibitem[\protect\astroncite{Uhlenbeck and Ornstein}{1930}]{Uhl30}
Uhlenbeck, G.~E. and Ornstein, L.~S. (1930).
\newblock On the theory of the brownian motion.
\newblock {\em Physical Review}, 36:823--841.

\bibitem[\protect\astroncite{Wascher et~al.}{2024}]{Wascher2024IDSA}
Wascher, M., Schnell, P., KhudaBukhsh, W.~R., Quam, M., Tien, J., and
  Rempa{\l}a, G.~A. (2024).
\newblock Estimating disease transmission in a closed population under repeated
  testing.
\newblock {\em Journal of the Royal Statistical Society: Series C}.

\bibitem[\protect\astroncite{Zhang et~al.}{2022}]{Zha22}
Zhang, M.~M., Williamson, S.~A., and P{\'e}rez-Cruz, F. (2022).
\newblock Accelerated parallel non-conjugate sampling for bayesian
  non-parametric models.
\newblock {\em Statistics and Computing}, 32(3):1--25.

\end{thebibliography}

\newpage
\appendix

\section{Additional details on the split-merge algorithm} \label{appendix:main_algorithm}

	In this section we provide further details on Algorithm~\ref{algo:main_algorithm}, which can be helpful for future implementations of the model discussed in the manuscript. The split-merge strategy we used in the manuscript is based on the preliminary studies of~\citet{JainNeal2004} and~\cite{JainNeal2007}, where we make use of the some results of the latter for non-conjugate distributions. Let us recall first the general specification of the model, with 
	\begin{equation}\label{mod:spec_app}
		\begin{split}
			\bm y_i \mid \rho_i, \bm \theta_i^* &\sim \prod_{j=1}^{m_i} \prod_{t = t_{i,j}^-}^{t_{i,j}^+} \Law(y_{i,t}\mid y_{i,t-1}, \theta_{i,j}^*),\quad i = 1, \dots, n,\\
			\rho_i \mid \tilde p(\rho)&\simiid \tilde p(\rho), \quad i = 1, \dots, n,\\
			(\pi_1,\dots,\pi_{2^{T-1}}) &\sim \textsc{Dir}(\alpha_1, \dots, \alpha_{2^{T-1}}),\\
			\theta_{i,j}^*&\simiid P_0(\theta),\quad j = 1, \dots, m_i, \; i = 1, \dots, n,
		\end{split}
	\end{equation}
	whereas 
	\begin{equation}\label{eq:disc_prior_app}
		\tilde p(\rho) = \sum_{r=1}^{2^{T-1}}\pi_r\delta_{ \tilde \rho_r}(\rho),
	\end{equation}
	$\{\tilde\rho_1, \dots, \tilde\rho_{2^{T-1}}\}$ denote all the possible orders of $T$ elements, and $(\pi_1, \dots, \pi_{2^{T-1}})$ is a vector of probabilities taking values in the $(2^{T-1}-1)$-dimensional simplex space. 
	
	Before discussing the sampling step, we introduce some quantities which will be needed in the derivation of the acceptance rates. We first note that the distribution of a sequence of elements $\mathcal R = \{\rho_1, \dots, \rho_n\}$ can be decomposed in the distribution of the partition structure $\lambda$ times the distribution of the unique elements $\mathcal R^* = \{\rho_1^*, \dots, \rho_k^*\}$, with $k \leq n$, and 
	\[
	\Law(\mathcal R) = \Law(\lambda) \Law(\mathcal R^*). 
	\]
	The first term is given by the EPPF of a Dirichlet-categorical model, with 
	\[
	\Law(\lambda) = \frac{\Gamma(\alpha^{+})}{\Gamma(\alpha^{+} + n)} \prod_{r=1}^{k} \frac{\Gamma(\alpha_r + n_r)}{\Gamma(\alpha_r)},
	\]
	with $\lambda = \{B_1, \dots, B_k\}$ and $n_j = \lvert B_j \rvert$. The second term is the probability associated with a specific set of unique values. As far as all the values have the same prior guess, the joint distribution of a specific subset is given by the law of a uniform sampling without replacement. Hence, we have 
	\begin{equation*}
		\Law(\mathcal R^*) = 1 - \prod_{j=0}^{k-1} \left(1-\frac{1}{2^{T-1} - j}\right),
	\end{equation*}
	where $k$ is the number of groups. Finally, we denote by 
	\[
	q(\lambda^{(N)}, \mathcal R^{(N)} \mid \lambda^{(O)}, \mathcal R^{(O)}, \mathcal Y)
	\]
	the proposal from a previous state $(\lambda^{(O)}, \mathcal R^{(O)})$ to a new state $(\lambda^{(N)}, \mathcal R^{(N)})$, where we emphasize the dependence on $\mathcal Y$ of the mixture strategy as proposal distribution for the new orders. Further, we have that the proposal factorises in two terms
	\begin{equation} \label{eq:prior_split_details}
		q(\lambda^{(N)}, \mathcal{R}^{(N)} \mid \lambda^{(O)}, \mathcal{R}^{(O)}, \mathcal Y)  
		= q(\lambda^{(N)} \mid \lambda^{(O)}) q(\mathcal{R}^{(N)} \mid  \mathcal{Y}), 
	\end{equation}
	whereas the first term in the left side of the previous equation is described in \cite{JainNeal2004} and does not depend on $\mathcal Y$, while the second term is a product of independent proposals following the mixture of posteriors discussed in the manuscript, i.e. 
	\begin{equation} \label{eq:proposal_app}
		\psi(\rho \mid \mathcal Y) = \sum_{i=1}^n \frac{1}{n} \Law(\rho \mid \bm y_i).
	\end{equation} 
	We can now describe a detailed single step of the split-merge strategy for clustering multiple time series. Suppose we are performing the generic $m$ step of the algorithm to cluster the observed data $\mathcal Y$, assuming we are on a current state of the chain, with $\lambda^{(m-1)} = \{B_1^{(m-1)},\dots, B_k^{(m-1)} \}$ being the latent partition of the data in $k$ groups at the current state and $\mathcal R^{*(m-1)} = \{ \rho_1^{*(m-1)}, \dots, \rho_{k}^{*(m-1)} \}$ denotes the unique values of the latent orders out of $\mathcal R^{(m-1)} = \{ \rho_1^{(m-1)}, \dots, \rho_{n}^{(m-1)} \}$, associated with each cluster in $\lambda^{(m-1)}$. 
	
	We initialize the proposed values $\lambda^{(N)}$ and $\mathcal R^{(N)}$ by setting the latent partition and the unique values of the random orders equal to the previous state values, i.e. we set $\lambda^{(N)} = \lambda^{(m-1)}$ and $\mathcal R^{*(N)} = \mathcal R^{*(m-1)}$. The split-merge step starts by sampling two indices $i \neq \ell$ from $\{1, \dots, n\}$. Then, one of the following cases can occur. 
	\begin{enumerate}
		\item[(a)] If the observations $i$ and $\ell$ belong to the same cluster, e.g. to the generic $s$ block $B_s^{(m-1)}$ of $\lambda^{(m-1)}$, then we perform a split step. 
		\begin{itemize}
			\item[-] We update the proposed partition $\lambda^{(N)}$ according to the split step: we assign $i$ to $B_s^{(N)}$ and $\ell$ to a new block $B_{k+1}^{(N)}$. We then randomly assign all the remaining observations in $B_s^{(m-1)} \setminus \{i, \ell\}$ to $B_S^{(N)}$  or $B_{k+1}^{(N)}$.
			\item[-] We update the proposed unique values of the random orders $\mathcal R^{*(N)}$, whereas we first increase the dimension of $\mathcal R^{*(N)}$ to $k+1$ and then the elements $s$ and $k + 1$ are sampled from the mixture of posterior distributions in~\eqref{eq:proposal_app}. 
			\item[-] Compute the acceptance rate
			\begin{multline*}
				\alpha(\lambda^{(N)}, \mathcal{R}^{*(N)} \mid \lambda^{(m-1)}, \mathcal{R}^{*(m-1)}) 
				= \min \left\{ 1, \frac{q(\lambda^{(m-1)}, \mathcal{R}^{*(m-1)} \mid \lambda^{(N)}, \mathcal{R}^{*(N)})}{ q(\lambda^{(N)}, \mathcal{R}^{*(N)} \mid \lambda^{(m-1)}, \mathcal{R}^{*(m-1)})} \right. \\ \times \left.  \frac{\Law(\lambda^{(N)}) \Law( \mathcal{R}^{*(N)})}{\Law(\lambda^{(m-1)}) \Law(\mathcal{R}^{*(m-1)})} \frac{\Law(\mathcal Y \mid \lambda^{(N)}, \mathcal{R}^{*(N)})}{L(\mathcal Y \mid \lambda^{(m-1)}, \mathcal{R}^{*(m-1)})} \right\}.
			\end{multline*} 
			Thanks to the factorization in~\eqref{eq:prior_split_details}, the first ratio in the acceptance rate becomes 
			\[
			\frac{q(\lambda^{(m-1)} \mid \lambda^{(N)})}{q(\lambda^{(N)} \mid \lambda^{(m-1)})} \frac{q(\mathcal{R}^{(m-1)} \mid  \mathcal{Y})}{ q(\mathcal{R}^{(N)} \mid  \mathcal{Y})} = 2^{|B_s^{(m - 1)}| - 2}\frac{\psi(\rho_s^{*(m-1)} \mid \mathcal{Y})}{\psi(\rho_s^{*(N)} \mid \mathcal{Y})\psi(\rho_{k+1}^{*(N)} \mid \mathcal{Y})}
			\]
			where the first term follows the derivation in~\citet{JainNeal2004} while the second is given by the mixture of posteriors proposal distributions. For the remaining terms, the followings hold
			\[
			\begin{split}
				\frac{\Law(\lambda^{(N)})}{\Law(\lambda^{(m-1)})} &= \frac{\Gamma\big(\alpha + n_s^{(N)}\big)\Gamma\big(\alpha + n_{k+1}^{(N)}\big)}{\Gamma(\alpha)\Gamma\big(\alpha + n_s^{(m-1)}\big)}\\
				\frac{\Law(\mathcal R^{*(N)})}{\Law(\mathcal R^{*(m-1)})} &= \frac{1 - \prod_{j=0}^{k} \left(1-\frac{1}{2^{T-1} - j}\right)}{1 - \prod_{j=0}^{k-1} \left(1-\frac{1}{2^{T-1} - j}\right)}\\
				\frac{\Law(\mathcal Y \mid \lambda^{(N)}, \mathcal{R}^{*(N)})}{\Law(\mathcal Y \mid \lambda^{(m-1)}, \mathcal{R}^{*(m-1)})} &= \frac{ \prod_{j \in B_{s}^{(N)}} \mathcal L\big(\bm y_{j} \mid \rho_s^{*(N)}\big) \prod_{j \in B_{k+1}^{(N)}}  \mathcal L\big(\bm y_{j} \mid \rho_{k+1}^{*(N)}\big) }{\prod_{j \in B_{s}^{(m-1)}} \mathcal L\big(\bm y_{j} \mid \rho_s^{*(m-1)}\big) }
			\end{split}
			\]
			where $n_s^{(N)}$ and $n_s^{(m-1)}$ denote the size of the $s$th block in the proposed partition and the previous state, respectively, and the latter equation follows the restricted Gibbs sampling strategy discussed in~\cite{JainNeal2004}. 
		\end{itemize}
		\item[(b)] If observations $i$ and $\ell$ belong to different clusters, e.g. $i \in B_s^{(m-1)}$ and $\ell \in B_r^{(m-1)}$ with $s < r$, then we perform a merge step. 
		\begin{itemize}
			\item[-] We update the proposed partition $\lambda^{(N)}$ according to the merge step: we assign all the observations in $B_r^{(m-1)}$ to $B_s^{(m-1)}$ and we destroy the block $B_r^{(m-1)}$. 
			\item[-] We update the proposed state for the unique values of the random orders $\mathcal R^{*(N)}$ by destroying the $r$th element and by sampling the $s$th element from the mixture of posterior distributions in~\eqref{eq:proposal_app}. 
			\item[-] Compute the acceptance rate
			\begin{multline*}
				\alpha(\lambda^{(N)}, \mathcal{R}^{*(N)} \mid \lambda^{(m-1)}, \mathcal{R}^{*(m-1)}) 
				= \min \left\{ 1, \frac{q(\lambda^{(m-1)}, \mathcal{R}^{*(m-1)} \mid \lambda^{(N)}, \mathcal{R}^{*(N)})}{ q(\lambda^{(N)}, \mathcal{R}^{*(N)} \mid \lambda^{(m-1)}, \mathcal{R}^{*(m-1)})} \right. \\ \times \left.  \frac{\Law(\lambda^{(N)}) \Law( \mathcal{R}^{*(N)})}{\Law(\lambda^{(m-1)}) \Law(\mathcal{R}^{*(m-1)})} \frac{\Law(\mathcal Y \mid \lambda^{(N)}, \mathcal{R}^{*(N)})}{L(\mathcal Y \mid \lambda^{(m-1)}, \mathcal{R}^{*(m-1)})} \right\}.
			\end{multline*} 
			Similarly to point (a), we have
			\[
			\frac{q(\lambda^{(m-1)} \mid \lambda^{(N)})}{q(\lambda^{(N)} \mid \lambda^{(m-1)})} \frac{q(\mathcal{R}^{(m-1)} \mid  \mathcal{Y})}{ q(\mathcal{R}^{(N)} \mid  \mathcal{Y})} = \left(\cfrac{1}{2} \right)^{|B_s^{(m - 1)}| + |B_r^{(m - 1)}| - 2} \frac{\psi(\rho_s^{*(m-1)} \mid \mathcal{Y})\psi(\rho_{r}^{*(m-1)} \mid \mathcal{Y})}{\psi(\rho_s^{*(N)} \mid \mathcal{Y})}.
			\]
			Further, for the remaining terms the followings hold
			\[
			\begin{split}
				\frac{\Law(\lambda^{(N)})}{\Law(\lambda^{(m-1)})} &= \frac{\Gamma(\alpha)\Gamma\big(\alpha + n_s^{(N)}\big)}{\Gamma\big(\alpha + n_s^{(m-1)}\big)\Gamma\big(\alpha + n_{r}^{()m-1}\big)},\\
				\frac{\Law(\mathcal R^{*(N)})}{\Law(\mathcal R^{*(m-1)})} &= \frac{1 - \prod_{j=0}^{k-1} \left(1-\frac{1}{2^{T-1} - j}\right)}{1 - \prod_{j=0}^{k} \left(1-\frac{1}{2^{T-1} - j}\right)},\\
				\frac{\Law(\mathcal Y \mid \lambda^{(N)}, \mathcal{R}^{*(N)})}{\Law(\mathcal Y \mid \lambda^{(m-1)}, \mathcal{R}^{*(m-1)})} &= \frac{\prod_{j \in B_{s}^{(N)}} \mathcal L\big(\bm y_{j} \mid \rho_s^{*(N)}\big) }{ \prod_{j \in B_{s}^{(m-1)}} \mathcal L\big(\bm y_{j} \mid \rho_s^{*(m-1)}\big) \prod_{j \in B_{r}^{(m-1)}}  \mathcal L\big(\bm y_{j} \mid \rho_{k+1}^{*(m-1)}\big) }.
			\end{split}
			\]
		\end{itemize}
	\end{enumerate}
	
	Finally, we perform a Metropolis--Hasting step by sampling a uniform random variable $U \sim Unif(0,1)$. If $U < \alpha(\lambda^{(N)}, \mathcal{R}^{*(N)} \mid \lambda^{(m-1)}, \mathcal{R}^{*(m-1)})$ we accept the proposed values $\lambda^{(N)}$ and $\mathcal R^{*(N)}$ as current state of the chain, setting $(\lambda^{(m)}, \mathcal R^{*(m)})$ equal to $(\lambda^{(N)}, \mathcal R^{*(N)})$, otherwise if $U > \alpha(\lambda^{(N)}, \mathcal{R}^{*(N)} \mid \lambda^{(m-1)}, \mathcal{R}^{*(m-1)})$ we stay on the previous state of the chain and we set $(\lambda^{(m)}, \mathcal R^{*(m)})$ equal to $(\lambda^{(m-1)}, \mathcal R^{*(m-1)})$.

	\section{Additional details on the SIR model}\label{app:epi}

	The scaled stochastic process $\bm \eX$ is a time-inhomogeneous continuous-time Markov process generated by \cite[Chapter 1]{Gikhman2004StochPro2} (see also \cite{Applebaum2019Semigroups,EngelNagel2000OneParameter})
	\begin{align}
		\begin{aligned}
			\eA_t g(x) &{}=  \varepsilon^{-1} \beta(t)  x_1 x_2 (g(x_1 -\varepsilon, x_2+\varepsilon, x_3) - g(x)) 
			+ \varepsilon^{-1} \xi(t) x_2 (g(x_1, x_2-\varepsilon, x_3+\varepsilon) -  g(x)), 
		\end{aligned}
		\label{eq:generator}
	\end{align}
	for bounded, continuous functions $g:\mathbb{R}_{+}^3 \mapsto \R$. The trajectories of the scaled process $\eX$ can be described as solutions to the following stochastic differential equations (written in the integral form)
	\begin{align}
		\begin{aligned}
			\eX_S(t) &{}= \eX_S(0) - \varepsilon \int_0^t \int_0^\infty \indicator{[0, \varepsilon^{-1}  \beta(u) \eX_S(u_-) \eX_I(u_-) ]}{v} Q_1(\D u, \D v) , \\
			\eX_I(t) &{}= \eX_I(0) + \varepsilon \int_0^t \int_0^\infty \indicator{[0, \varepsilon^{-1}  \beta(u) \eX_S(u_-) \eX_I(u_-) ]}{v} Q_1(\D u, \D v) \\
			&{}\quad \quad - \varepsilon\int_0^t \int_0^\infty \indicator{[0, \varepsilon^{-1}  \xi(u) \eX_I(u_-) ]}{v} Q_2(\D u, \D v), \\
			\eX_R(t) &{}= \eX_R(0) + \varepsilon \int_0^t \int_0^\infty \indicator{[0, \varepsilon^{-1}  \xi(u) \eX_I(u_-) ]}{v} Q_2(\D u, \D v), 
		\end{aligned}
		\label{eq:scaled_trajectory_equations}
	\end{align}
	where $Q_1, Q_2$ are independent Poisson random measures on $\R^2$ with intensity measure $\D u \times \D v$ where $\D u$, and $\D v$ are Lebesgue measures on $\R$. We have used the notation $\indicator{A}{x}$ to denote the indicator function of a set $A$, which takes the value one when $x\in A$, and zero otherwise. The process $X$ (or the scaled process $\eX$) can simulated as a pure jump process using standard techniques. An alternative individual-based approach is provided by what is known as the \emph{Sellke construction} \cite{AnderssonBritton2000}. Let $H_\varepsilon(t)$ denote the cumulative infection pressure upto time $t$, where 
	\begin{align}
		H_\varepsilon(t) = \int_0^t \varepsilon \beta(u) X_I(u) \D u= \int_0^t  \beta(u) \eX_I(u) \D u. 
	\end{align}
	According to the Sellke construction, each susceptible individual is assigned an exponentially distributed (with mean one) threshold $E$, and gets infected at time $t$ as soon as  $E \le H_\varepsilon(t)$. If such a time never arrives, \emph{i.e.}, if $E > \lim_{t\to\infty}H_\varepsilon(t) $ (the limit is finite almost surely), then the individual escapes infection. That is, conditional on the history of the process up to time $t$, the probability that a randomly chosen susceptible individual is still susceptible at time $t$, \emph{i.e.}, has not been infected by time $t$, is $\exp\left(- H_\varepsilon(t)\right)$. This individual-based perspective is at the heart of the Dynamical Survival Analysis (DSA) \cite{KhudaBukhsh2020DSA,DiLauro2022NonMarkov} approach  to parameter inference for infectious disease epidemiology based on a random sample of infection times (and recovery times, if available) as opposed to standard methods that require (or impute when full data are not available) population-level trajectories.

	To study the large population limit of the scaled stochastic process $\eX$, let us rewrite the trajectory in \eqref{eq:scaled_trajectory_equations} as 
	\begin{align}
		\begin{aligned}
			\eX_S(t) &{}= \eX_S(0) - \eM_S(t) -   \int_0^t  \beta(u) \eX_S(u_-) \eX_I(u_-)  \D u, \\
			\eX_I(t) &{}= \eX_I(0) + \eM_I(t) + \int_0^t  \beta(u) \eX_S(u_-) \eX_I(u_-) \D u - \int_0^t \xi(u) \eX_I(u_-) \D u, \\
			\eX_R(t) &{}= \eX_R(0) + \eM_R(t) +   \int_0^t \xi(u) \eX_I(u_-)\D u,
		\end{aligned}
		\nonumber 
	\end{align}
	where the stochastic processes $\eM_S, \eM_I$, and $\eM_R$ are square-integrable zero-mean martingales defined as 
	\begin{align}
		\begin{aligned}
			\eM_S(t) &{}= \varepsilon \int_0^t \int_0^\infty \indicator{[0, \varepsilon^{-1}  \beta(u) \eX_S(u-) \eX_I(u-) ]}{v} \bar{Q}_1(\D u, \D v), \\
			\eM_I(t) &{}= \varepsilon \int_0^t \int_0^\infty \indicator{[0, \varepsilon^{-1}  \beta(u) \eX_S(u-) \eX_I(u-) ]}{v} \bar{Q}_1(\D u, \D v) - \varepsilon \int_0^t \int_0^\infty \indicator{[0, \varepsilon^{-1}  \xi(u) \eX_I(u-) ]}{v} \bar{Q}_2(\D u, \D v), \\
			\eM_R(t) &{}= \varepsilon \int_0^t \int_0^\infty \indicator{[0, \varepsilon^{-1}  \xi(u) \eX_I(u-) ]}{v} \bar{Q}_2(\D u, \D v),
		\end{aligned}
		\nonumber 
	\end{align}
	where $\bar{Q}_1$ and $\bar{Q}_2$ are the compensated Poisson random measures (corresponding to $Q_1$ and $Q_2$ respectively).  It is easy to verify that the quadratic variations of the above martingales all converge to zero in probability as $\varepsilon \to 0$, which in turn imply that the martingales themselves converge to the deterministic function taking the value zero at all times. Consider the system of Ordinary Differential Equations (ODEs) in \eqref{eq:meanfield_SIR_ODE}
	with the initial condition $S(0)=1, I(0) = I_0 $, and $R(0)=0$. Let $\bar{X} = (S, I , R)$.  Then, note that $\eX(0) \to (1, I_0, 0)$ as $\varepsilon\to 0$ by design, and as a consequence of the Functional Law of Large Numbers (FLLN), we can prove that 
	\begin{align}
		\lim_{\varepsilon \to 0} \sup_{t \le T} \norm{\eX(t) - \bar{X}(t) }{\infty} \to 0 
	\end{align}
	in probability where $\norm{(x_1, x_2, x_3)}{\infty} = \max\{|{x_1}|, |x_2|, |x_3|\}$. In fact, the above convergence holds almost surely since we have assumed the functions $\beta(t)$ and $\xi(t)$ are bounded. The proof follows by standard arguments using Gr\"onwall's inequality and some maximal estimates. See, for example, \cite[Theorem 3.1]{KhudaBukhsh2024SIR_SEIR}, where it is also discussed how an SIR model with time-varying infection and recovery rates such as the model we use in this paper can be used to approximate a compartmental Susceptible-Exposed-Infected-Recovered (SEIR) model.

	\section{Synthetic illustration with time series} \label{subsec:application_time_series}
	
	Before extending our model strategy to more complicated scenarios, we first investigate the performance of Algorithm~\ref{algo:main_algorithm} with a synthetic study. We consider a case where data are real-valued time series, for which structural changes occur at certain times. The generic $j$th block of the $i$th time series is generated according to the following process
	\begin{equation} \label{eq:simulation}
		y_{i,t} = 
		\begin{cases}
			\mu_{i,t} + N(0,\eta_{i,t}), & \text{if } t = 1,\\ 
			\gamma \, y_{i,t-1} + (1 - \gamma) \, \mu_{i,t} + N(0, (1 - \gamma^2) \eta_{i,t}),  & \text{otherwise,}
		\end{cases}
	\end{equation}
	for $j = 1, \dots, k_i$, where $N(a, b)$ denotes the Gaussian distribution with expectation $a$ and variance $b$. We consider a scenario with three groups of time series, whereas each series is observed $T = 300$ times divided in $k_i = 3$ distinct regimes, $i=1, \dots, 10$, and series belonging to the same cluster have change points at the same time but different local parameters. For all the time series, we fix $\gamma = 0.1$ and set the unique values of the parameter $\mu_{i,j}^*$s, out of $\mu_{i,1}, \dots, \mu_{i,T}$ and $\eta_{i,j}^*$s, out of $\eta_{i,1}, \dots, \eta_{i,T}$, according to the scheme in Table \ref{table:sim_setup}. An example of the sampled data is shown in Figure \ref{fig:simulations_TS_and_psm}.
	
	\begin{table}[h] 
		\centering
		
		\begin{tabular}{cccc}
			\midrule 
			$i$ & $\{\mu_{i,1}^*, \dots, \mu_{i,k_i}^*\}$ & $\{\eta_{i,1}^*, \dots, \eta_{i,k_i}^*\}$ & $\{\lvert A_{i,1}\rvert, \dots, \lvert A_{i,k_i}\rvert\}$  \\ 
			\midrule 
			1 & \{0.5,0.85,0.5,0.75,1\} & \{0.1,0.12,0.14,0.13,0.15\} & \multirow{4}{*}{\{50,100,45,55,50\}} \\
			2  & \{0.15,0.75,0.25,0,0.25\} & \{0.12,0.15,0.12,0.14,0.13\} &  \\
			3  & \{0.25,0,0.15,0.15,0.3\} & \{0.1,0.12,0.2,0.12,0.14\} &  \\ 
			4 &  \{0.75,0.4,0.8,0.8,0.4\} & \{0.1,0.12,0.09,0.24,0.15\} &  \\
			\midrule 
			5 & \{0,-0.15,0.15,0.3,0.1,0.3\} &\{0.12,0.13,0.1,0.13,0.14,0.12\}   &  \multirow{3}{*}{\{40,50,45,45,30,90\}} \\
			6 & \{0.5,0,-0.5,0,0.2,0\}  &  \{0.1,0.24,0.14,0.15,0.12,0.13\} & \\
			7 & \{0,0.2,0.4,0.25,-0.1,0.15\}  & \{0.16,0.15,0.1,0.13,0.14,0.12\}    &   \\
			\midrule 
			8 & \{0,-0.25,0,0.25,-0.25,0.1\} & \{0.14,0.13,0.17,0.12,0.14,0.12\} &  \multirow{3}{*}{\{75,75,30,20,75,25\}}   \\
			9  & \{0.25,0.25,-0.2,0.1,0.3,0\} & \{0.12,0.22,0.15,0.14,0.17,0.19\}  & \\
			10 & \{0,-0.25,0,0.25,0,-0.25\}  & \{0.12,0.13,0.15,0.12,0.15,0.18\}  & \\ 
			\midrule 
		\end{tabular}
		\caption{Parameters of the data generating processes for the the synthetic study with real-valued time series. Left to right: observation index, local trend parameters, local dispersion parameters, and true latent order shared among observations in the same cluster.}\label{table:sim_setup}
	\end{table}
	
	The data are analysed by considering a model as in~\eqref{mod:spec}, where the likelihood term factorises in real-valued models. In particular, we assume that within each block the data follow an univariate Ornstein-Uhlenbeck process \citep{Uhl30}. The process defined as the solution to the following stochastic differential equation
	\begin{equation}
		dY_t = -\alpha(Y_t - \mu)dt + \sqrt{\frac{2\alpha}{\eta}}dW_t,
	\end{equation}
	where $\mu \in \mathbb R$, $\alpha, \eta > 0$, $\{W_t\}_{t \geq 0}$ is the Wiener process, and $\alpha$ is the parameter tuning the dependence over time. Here, the generic parameter indexing locally the time series is given by $\theta_{i,t} = (\mu_{i,t}, \eta_{i,t})$, while the parameter $\alpha$ (or equivalently $\gamma = e^{-\alpha}$) is assumed to be fixed and shared for all the time series. By setting $\mu \mid \eta \sim \text{Normal}(0, (c\eta)^{-1})$ and $\eta \sim \text{Gamma}(a,b)$, \cite{Mar14} have shown that the marginal distribution of the data becomes

	\begin{align}
		\label{eq:integrated_likelihood_ts}
		\begin{split}
			\mathcal{M}(\{y_{i,t}:t \in A_{i,j}\} \mid \gamma) &= \frac{(2b(1-\gamma^2))^a \Gamma(n_j/2 + a)}{\pi^{n_j/2}\Gamma(a)}\left(\frac{c(1+\gamma)(1-\gamma^2)}{c+n_j-\gamma(n_j-c-2)}\right)^{1/2} \times \\ &\left( \bm y_j^{\intercal} \bm S_j \bm y_j - \frac{(1-\gamma) (\sum_{i=1}^{n_j}y_{j,i} - \gamma \sum_{i=2}^{n_j-1}y_{j,i})^2}{c+n_j-\gamma(n_j-c-2)} + 2b(1-\gamma^2) \right)^{-(n_j/2 + a)},
		\end{split}
	\end{align}
	
	where $\bm y_j = \left(y_{i,t_{i,j}^-}, \dots, y_{i,t_{i,j}^+}\right)$ and $\bm S_j$ is an $n_j \times n_j$ matrix of the form 
	
	\begin{center}
		$\bm S_j = \begin{bmatrix} 
			1 & -\gamma & 0 & \dots  & 0\\
			-\gamma & 1 + \gamma^2 & -\gamma & \dots & 0 \\
			0 & -\gamma & 1 + \gamma^2 & \dots & 0 \\
			\vdots & \vdots & \vdots & \ddots & \vdots \\
			0 & 0 & 0 & \dots & 1
		\end{bmatrix}$.     
	\end{center}
	
	Posterior inference is performed exploiting Algorithm~\ref{algo:main_algorithm}. We consider different scenarios corresponding to different specification of such algorithm. In particular, we consider different values of $\mathrm B \in \{1\,000, 10\,000, 100\,000\}$, which corresponds to the number of points used to evaluate via importance sampling the normalization constants in Equation~\ref{eq:proposal}, a parameter that tunes the accuracy of those constants.
	The second quantity which varies across the scenarios is the depth of the proposal, i.e. the number of sampling steps $\mathrm L$ that we are performing to propose a candidate from Equation~\ref{eq:proposal}. Specifically, while proposing a candidate random order, we are randomly selecting one of the $n$ dimensions in~\eqref{eq:proposal}, then we sample from the corresponding posterior distribution. For the latter step, we initialise randomly the latent order and we then perform $L$ split-merge step \citep{Mar14} to update that value. Each scenario considered in the study is replicated $50$ times. Data are standardized before running the model, so that all the observations contributes equally to the computation of the likelihood. We assign a Laplace's prior for the parameters of the Dirichlet distribution in Equation~\ref{eq:mod_int}, so $\alpha_r = \alpha = 1$ for every $r \in \{1,\dots,2^{T-1}\}$. Following the approach by \cite{Mar14}, the likelihood of a time series $y_i$ is given by \eqref{eq:integrated_likelihood_ts} where $\mu$ and $\eta$ have been marginalised after setting respectively a normal and a gamma prior distributions. According to the notation introduced beforehand, we set as parameters for the gamma  $a = b = 1$ and $c=0.1$ for the normal distribution. Finally, we set the autocorrelation $\gamma = 0.1$. For the split-merge procedures that compute the marginal change points we set the probability of performing a split to $q=0.5$.
	We run each simulation for $5\,000$ iterations of which $2\,000$ are discarded as burn-in period. We then obtain the point estimate of the latent partition $\hat\lambda$ as the one among those visited by the algorithm that minimizes the expected Binder loss function \citep{Binder1978}.

	\begin{table}[!h] 
		\centering
		\begin{tabular}{cccccccccccc}
			\toprule
			$\mathrm B$ & $\mathrm L$ &  $\mathrm{BI}(\hat \lambda, \lambda_0)$ & $\mathrm B$ & $\mathrm L$ & ${\mathrm{BI}}(\hat \lambda, \lambda_0)$ & $\mathrm B$ & $\mathrm L$ & ${\mathrm{BI}}(\hat \lambda, \lambda_0)$ \\ \midrule
			\multirow{3}{*}{1\,000}  
			& 1 & 0.076 & \multirow{3}{*}{10\,000} & 1 & 0.028  &\multirow{3}{*}{100\,000} & 1 & 0.038   \\
			& 25  & 0.027 & & 25 & 0.018 & & 25 & 0.035   \\
			& 100 & 0.033  &  & 100 & 0.017 & & 100 & 0.038    \\
			\bottomrule
		\end{tabular}
		
		\caption{Posterior summaries of the time series simulation study. We consider different scenarios, where $\mathrm B$ denotes the accuracy of the normalization constants and $\mathrm L$ represents the proposal depth. The table reports the Binder loss between the point estimate and the true latent partition of the data ($\mathrm{BI}$). Results are averaged over $50$ replicates. 
		}
		\label{table:results_simulation_ts}
		
	\end{table}
	
	Table \ref{table:results_simulation_ts} shows the summaries of interest of the time series simulation study, namely the Binder loss function between the true and the estimated partition of the data, averaged over the $50$ replicates. The algorithm performs well consistently with all configurations. Both the accuracy of the normalization constants $\mathrm B$ and the proposal depth $\mathrm L$ do not seem to affect systematically the performance of Algorithm~\ref{algo:main_algorithm}. Intuitively, large $\mathrm B$ and $\mathrm L$ should be preferred, as they increase the precision of the proposal and of its sampling step. However, small values of $\mathrm L$ strongly reduce the computational time needed to perform the sampling. 
	
	
	\section{Additional figures, tables and algorithms}\label{app:additional_fig_and_algo}
	
\begin{figure}[!h]
		\centering
		\includegraphics[scale = 0.82]{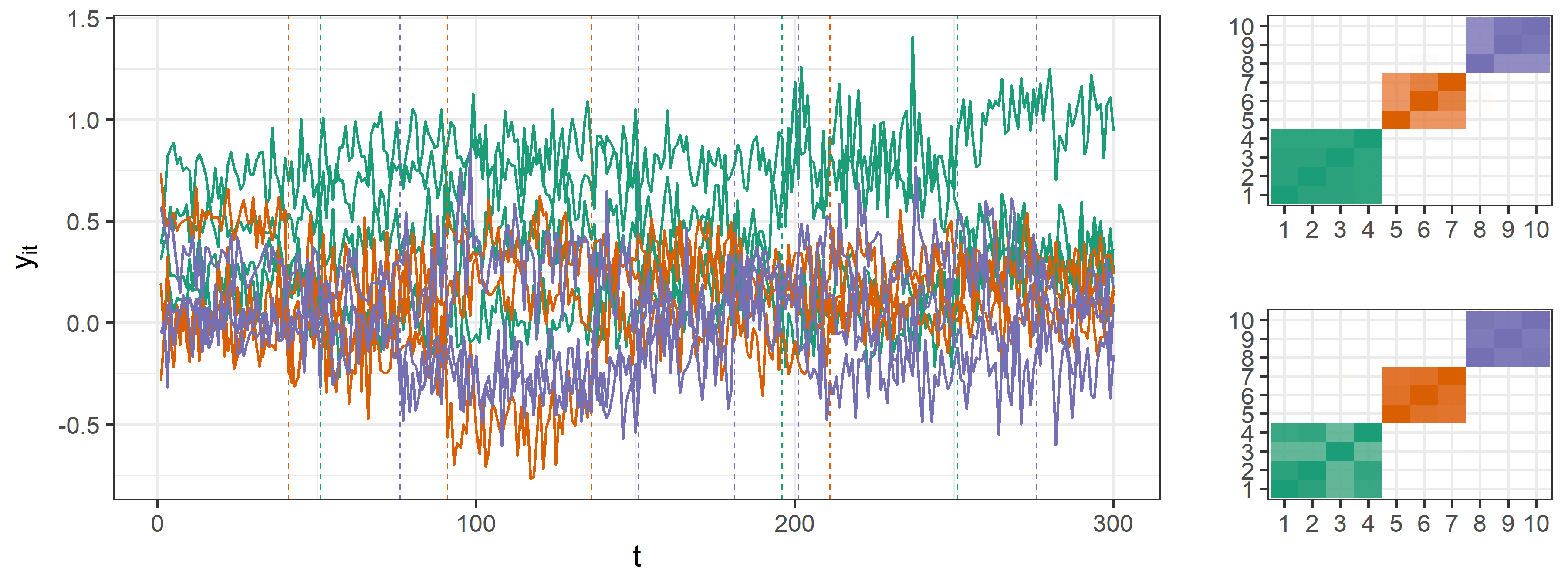}
		\caption{Left panel: one realization of simulated data according to the model in \eqref{eq:simulation} and parameters in Table \ref{table:sim_setup}. Right panels: posterior similarity matrix of one replication with $\mathrm B = 10\,000$, top with $\mathrm L = 1$ and bottom with $\mathrm L = 100$ respectively.
		} 
		\label{fig:simulations_TS_and_psm}
	\end{figure}

	In this section we report additional materials to support the manuscript, such as figures and algorithms. First, we show in Figure~\ref{fig:simulations_TS_and_psm} an example of simulated data for the time series synthetic study. The data generating process and observation-specific parameters are described in Section~\ref{subsec:application_time_series}. The figure shows in the left panel different time series, whereas each series is an observation characterized by its local parameters, but some of the series share the same structural change times. The right part of the figure shows two posterior similarity matrices, associated with two distinct estimates of the model, with the same accuracy of the normalization constant $\mathrm B = 10\,000$, but varying the depth of the proposal, top with $\mathrm L = 1$ and bottom with $\mathrm L = 100$. 
	
	Sampling from~\eqref{eq:proposal} is unfeasible due to the intractability of the normalization constant in each component of the mixture $\mathcal{L}(\rho \mid \bm y_i), \; i=1, \dots, n$. Normalising each component requires the computation of $\sum_{r = 1}^{2^{T-1}} \mathcal{L}(\tilde{\rho}_r \mid \bm y_i)$ for each observation $y_i$, where $\{ \tilde{\rho}_1, \dots \tilde{\rho}_{2^{T-1}}\}$ are possible orders of $T$ elements. Since the number of elements explodes as far as $T$ increases, evaluating the normalization constants becomes soon unfeasible. We then proceed by pre-computing these constants using an importance sampling strategy: the importance distribution follows a binomial-multinomial law, for which we first generate the number of blocks $k$ from a binomial distribution and then the frequencies from a multinomial distribution of dimension $k$. Then, we proceed with a standard importance re-weighting approach. Algorithm~\ref{algo:norm_const} shows an implementation of such strategy for a single normalization constant. 
	
	\begin{algorithm}[H]
		\DontPrintSemicolon
		\textbf{input} number of sampled orders $\mathrm B$ for the importance sampling (accuracy) and success probability $p \in (0,1)$ of the binomial distribution.\\
		\For{$r$ $\in$ $1,\dots,\mathrm B$} 
		{
			\vspace{3.5pt}
			\begin{enumerate}
				\item[a)] \textbf{sample} $k \sim \text{Binomial}(T-1, \, p)$.
				\item[b)] \textbf{sample} $\big(\lvert B_1^{(r)}\rvert, \dots, \lvert B_k^{(r)}\rvert\big)^\top \sim Multinom(T, \, ( 1/k, \dots, 1/k )^\top)$ and set $\rho^\dagger_r = \{B_1^{(r)}, \dots, B_k^{(r)}\}$.
				\item[d)] \textbf{compute} $\displaystyle{\phi(\rho^\dagger_r) = \binom{T}{k} \, p^k \, (1-p)^{T-k} \,  T! \,  \prod_{i = 1}^{k}\frac{1}{\big(k^{\lvert B_i^{(r)}\rvert}\big)\big(\lvert B_i^{(r)}\rvert!\big)}}$. 
			\end{enumerate}
		}
		\textbf{compute} the approximation of $\displaystyle{\sum_{r = 1}^{2^{T-1}} \mathcal{L}(\tilde{\rho}_r \mid \bm y_i) \simeq \frac{2^{T-1}}{\mathrm B} \sum_{r = 1}^{\mathrm B} \frac{1}{\phi(\rho^\dagger_r)} \mathcal{L}(\rho^\dagger_r \mid \bm y_i)}$.
		
		\textbf{end}
		\caption{\label{algo:norm_const} Computation of the $i$th normalization constant of~\eqref{eq:proposal}}
	\end{algorithm}
	
	Finally, Algorithm~\ref{algo:sellke_construction} shows the pseudo-code for producing a sample of infection times in a fixed time window, bounded by the last observational time $T$. Here we consider an algorithm that works with a functional infection rate $\beta(u)$, which in our case is obtained as step function starting from a vector of local infection rates $\bm \beta$. 
	
	
	\medskip
	\LinesNotNumbered
	\begin{algorithm}[H]
		\DontPrintSemicolon
		\textbf{input}{\, a functional ${\beta(u)}$ of time-varying infection rates and a value $\xi$ for the recovery rate, starting values for $S_0$, $I_0$ and $R_0$}.\\
		
		\textbf{denote}{\, with $(S_t,I_t,R_t)$ the vectors of susceptible, infected and recovered individuals at time $t$}.\\
		
		\textbf{set}{\, $t=0$ and $S_t = S_0$, $I_t = I_0$, $R_t = R_0$}.\\
		
		\While{$t < T$, $I_t > 0$ and $S_t > 0$} {
			
			\begin{enumerate}
				
				\item[a)] \textbf{sample} $E_1$ from $P(E_1>e_1) = \exp\left(-\frac{S_t I_t}{S_0}    \int_{t}^{t+e_1}\beta(u)du \right)$ and $E_2$ $\sim$ $\mathrm{Exp}\left( \xi I_t \right)$. \\
				
				\item[b)] \textbf{set} $t^* = \mathrm{min}(E_1, E_2)$. \\
				
				\begin{enumerate}
					\item[-] \textbf{if} $t^* = E_1$, then set $S_t = S_t - 1$, $I_t = I_t + 1$, $R_t = R_t$, and $\delta = 1$.
					\item[-] \textbf{else if} $t^* \neq E_1$, then set $S_t = S_t$, $I_t = I_t - 1$, $R_t = R_t + 1$, and $\delta = 0$.
				\end{enumerate}
				
				\item[c)] Set $t = t + t^*$. If $\delta = 1$ \textbf{then} save $t$ as new infection time. \\
			\end{enumerate}
		}
		\textbf{end}
		\caption{\label{algo:sellke_construction}Doob-Gillespie algorithm to sample infection times}
	\end{algorithm}

	\begin{table}[h]
		\centering
		\begin{tabular}{cccc}
			\midrule
			\multicolumn{1}{c}{$i$} & \multicolumn{1}{c}{$\{ \beta_1^*, \beta_2^*\}$}  & \multicolumn{1}{c}{$I_0$} & \multicolumn{1}{c}{$\{\lvert A_{i,1}\rvert, \lvert A_{i,2}\rvert\}$}   \\
			\midrule
			1 & \{0.211, 0.55\} & 23/$\mathrm{S0}$ & \multirow{4}{*}{\{110, 40\}}\\
			2 & \{0.221, 0.50\} & 23/$\mathrm{S0}$ &  \\
			3 & \{0.218, 0.54\} & 21/$\mathrm{S0}$ & \\
			4 & \{0.225, 0.51\} & 20/$\mathrm{S0}$ & \\ 
			\midrule
			5 & \{0.213, 0.52\} & 24/$\mathrm{S0}$ & \\
			6 & \{0.213, 0.51\} & 23/$\mathrm{S0}$ & \{90, 60\} \\
			7 & \{0.193, 0.57\} & 22/$\mathrm{S0}$ & \\
			\midrule
			8 & \{0.195, 0.54\} & 21/$\mathrm{S0}$  & \\
			9 & \{0.191, 0.53\}  & 20/$\mathrm{S0}$  & \{50, 100\} \\
			10 & \{0.189, 0.51\} & 24/$\mathrm{S0}$  & \\
			\midrule
		\end{tabular}
		\caption{Parameters of the data generating processes for the epidemiological synthetic study. Left to right: observation index, different local infection rates for each series, starting infection rate and true latent order shared among observations in the same cluster.} 
		\label{tab:sim_EPI_parameters}
	\end{table}

\end{document}